\documentclass[a4paper,11pt]{article}
\usepackage{jcappub} 
\usepackage{lineno}
\usepackage{amsmath,amssymb}
\usepackage{hyperref}
\usepackage{xcolor}
\usepackage{todonotes}
\usepackage{array}
\usepackage{subcaption}
\usepackage{graphicx}
\usepackage{dcolumn}
\usepackage{bm}
\usepackage{tensor}
\usepackage{amssymb}
\usepackage{slashed}
\usepackage{comment}
\usepackage{placeins}
\author[a]{Giorgio Laverda,}
\author[b,c,1]{Tomás Mendes\note{Corresponding author},}
\author[c]{Javier Rubio}
\affiliation[a]{PRISMA++ Cluster of Excellence \& Mainz Institute for Theoretical Physics, Johannes Gutenberg-Universität, Staudingerweg 9, 55099 Mainz, Germany}
\affiliation[b]{Centro de Astrofísica e Gravitação - CENTRA, Departamento de Física, Instituto Superior Técnico - IST, Universidade de Lisboa - UL, Av. Rovisco Pais 1, 1049-001 Lisboa, Portugal}
\affiliation[c]{Departamento de Física Teórica and Instituto de Física de Partículas y del Cosmos (IPARCOS-UCM), Universidad Complutense de Madrid, 28040 Madrid, Spain} 

\emailAdd{laverdag@uni-mainz.de, tomaslmendes@tecnico.ulisboa.pt, javier.rubio@ucm.es}
\abstract{We propose a novel gravitational mechanism for the non-thermal production of dark matter driven by curvature-induced tachyonic instabilities after inflation. Departing from the commonly studied non-minimal couplings to gravity, our framework considers a real spectator scalar field coupled quadratically to spacetime curvature invariants. We show that the rapid reorganization of spacetime curvature at the end of inflation can dynamically render the dark matter field tachyonic, triggering a short-lived phase of spontaneous symmetry breaking and explosive particle production. As a concrete and theoretically controlled example, we focus on the Gauss-Bonnet topological invariant. By combining analytical estimates with $3+1$ classical lattice simulations in the spectator field approximation, we track the out-of-equilibrium evolution of the system and compute the resulting dark matter abundance. We find that this purely gravitational mechanism can robustly reproduce the observed dark matter relic density over a wide range of masses and inflationary scales, providing also a simple fitting function that enables a lattice-independent application of our results.}

\begin{document}

\title{Tachyonic gravitational dark matter production after inflation}

\date{\today}

\maketitle
\flushbottom

\section{Introduction\protect}\label{sec:level1}

The existence of dark matter (DM) constitutes one of the most compelling pieces of evidence for physics beyond the Standard Model (SM) of particle physics. A wide range of observations, including galaxy rotation curves \cite{rubin1970}, gravitational lensing \cite{Clowe:2003tk}, large-scale structure measurements \cite{eBOSS:2020yzd}, and the anisotropies of the cosmic microwave background (CMB) \cite{Planck2018}, consistently point towards the presence of a non-luminous, matter-like component interacting predominantly through gravity. However, despite decades of experimental effort, the particle nature of DM remains unknown, and no conclusive evidence for particle DM candidates has been found. This includes, in particular, extensive searches in direct-detection experiments \cite{Schumann:2019eaa,XENON:2023cxc,LZ:2024zvo}, indirect-detection observations \cite{Gaskins:2016cha,Klasen:2015uma}, and collider probes \cite{Kahlhoefer:2017dnp,ATLAS:2016cfj,CMS:2017prz}. As a result, well-motivated and extensively studied DM candidates, such as Weakly Interacting Massive Particles (WIMPs) \cite{Steigman:1984ac,Bertone:2016nfn} or axions \cite{Preskill:1982cy,Marsh:2015xka,Adams:2022pbo}, remain experimentally elusive \cite{Bertone:2004pz,Arcadi:2017kky,Cirelli:2024ssz}.

The persistent lack of DM detection has increasingly motivated the exploration of alternative production and interaction mechanisms, ranging from macroscopic and non-thermal DM candidates, such as Q-balls, oscillons, boson stars, primordial black holes, or ultra-compact minihalos \cite{Ruffini:1969qy,Carr:1974nx,Coleman:1985ki,Colpi:1986ye,Kolb:1993zz,Navarro:1995iw,Kusenko:1997si,Braaten:2015eeu,Delos:2018ueo,Visinelli:2021uve,Carr:2024nlv,Amendola:2017xhl,Savastano:2019zpr,Goh:2023mau} to hidden sector particles \cite{Holdom:1985ag,Almeida:2018oid,Fabbrichesi:2020wbt,Cembranos:2024pvy} or fuzzy DM \cite{Hu:2000ke,Ferreira:2020fam}. A particularly appealing possibility is the purely gravitational production of DM, which does not require any direct coupling to the SM content and is therefore largely unconstrained by laboratory experiments. In this framework, DM particles can be generated through several mechanisms, including the Hawking evaporation of primordial black holes \cite{Green:1999yh,Khlopov:2004tn,Dai:2009hx,Fujita:2014hha,Allahverdi:2017sks,Lennon:2017tqq,Morrison:2018xla,Hooper:2019gtx,Chaudhuri:2020wjo,Masina:2020xhk,Baldes:2020nuv,Gondolo:2020uqv,Bernal:2020kse,Bernal:2020ili,Bernal:2020bjf,Cheek:2021odj,Cheek:2021cfe,Bernal:2021yyb,Bernal:2021bbv,Barman:2021ost,Barman:2022gjo,Bernal:2022oha,Cheek:2022dbx,Mazde:2022sdx,Cheek:2022mmy}, $s$-channel exchange of massless gravitons
\cite{Garny:2015sjg,Garny:2017kha,Tang:2017hvq,Bernal:2018qlk,Barman:2021ugy,Clery:2021bwz,Mambrini:2021zpp,Clery:2022wib,Barman:2022qgt,Ahmed:2022tfm,Haque:2023yra,Kaneta:2023uwi}, or non-perturbative effects in time-dependent backgrounds, including, in some cases, non-minimal couplings to gravity \cite{Ford:1986sy,Markkanen:2015xuw,Tang:2016vch,Ema:2018ucl,Hashiba:2018tbu,Fairbairn:2018bsw,Cembranos:2019qlm,Kainulainen:2022lzp,Kainulainen:2024etd,Chung:2001cb,Alonso-Alvarez:2018tus,Bernal:2020qyu,Laulumaa:2020pqi,Babichev:2020yeo,Lebedev:2022vwf,Barman:2023opy,Cembranos:2023urh,Cembranos:2023qph,Dorsch:2024nan,Cembranos:2025hvc, Davoudiasl:2024grq}. Under suitable conditions, all these scenarios can successfully account for a relic DM abundance consistent with observations.

In this work, we develop a novel class of gravitational DM production mechanisms that arise from curvature-induced tachyonic instabilities in the effective field theory (EFT) of gravity. A key observation underlying our analysis is that transitions between subsequent cosmological epochs with different equation-of-state parameters provide natural and efficient triggers for such dynamics. In particular, the transition from inflation to radiation domination (RD) is characterized by a sharp reorganization of curvature invariants, which can cause the effective mass of a non-minimally coupled spectator field to change sign. When this happens, the system undergoes a spontaneous symmetry-breaking driven entirely by gravitational effects. During this phase, the field experiences a tachyonic instability and a rapid amplification of quantum fluctuations, leading to efficient particle production. DM is thus generated as a consequence of the interplay between higher-curvature corrections and the expansion history of the Universe, without requiring direct couplings to the SM particles or finely tuned initial conditions.

Within this general framework, a wide class of quadratic curvature operators can in principle realize curvature-induced symmetry breaking. In this sense, the mechanism described here is not tied to a specific operator choice, but rather reflects a broader feature of gravitational EFTs coupled to spectator fields. For concreteness and theoretical control, however, we focus on the Euler or Gauss–Bonnet combination as a benchmark realization. When coupled to the spectator field, the Gauss–Bonnet invariant leads to a curvature contribution whose sign is fixed by the background equation of state, guaranteeing a positive effective field mass during inflation and a negative one during RD. At the same time, the Gauss–Bonnet term preserves second-order equations of motion for the metric and does not introduce additional propagating gravitational degrees of freedom. This allows us to isolate the curvature-driven instability as a property of the spectator sector evolution on a cosmological background, rather than as an artifact of modified gravity.

To assess the viability of our scenario as a source of the observed DM abundance, we combine analytical and numerical techniques. We first analyze the system in the homogeneous and non-homogeneous approximations, characterizing the conditions under which the curvature-induced instability develops and identifying the main parameter dependencies of the resulting DM relic density. Subsequently, we employ 3+1 classical lattice simulations to fully capture the evolution of the system through the symmetry-breaking phase and determine the resulting DM relic abundance with controlled accuracy. We find that curvature-induced gravitational production can robustly account for the observed DM density across a broad region of parameter space.

Beyond providing an alternative to conventional WIMP-like scenarios, this framework opens new and qualitatively different directions for DM model building rooted in gravitational EFTs. In this class of scenarios, spacetime curvature is not merely a passive background ingredient, but an active dynamical agent that can trigger symmetry breaking and particle production in otherwise secluded dark sectors. As a result, the properties and abundance of DM become directly linked to the expansion history of the Universe rather than to non-gravitational interactions with the visible sector. Moreover, the dynamics associated with the curvature-driven phase transition may leave observable imprints in cosmological observables, including stochastic gravitational-wave backgrounds, thereby providing potential avenues for testing the gravitational origin of DM with future observations.

This paper is organized as follows. In Section~\ref{sec:model} we introduce the EFT framework, motivating the Gauss–Bonnet combination as a concrete and theoretically controlled example. In Section~\ref{sec:analytics} we derive the associated equations of motion, quantize the spectator field, and solve approximately the resulting mode equations in the uniform approximation. This allows us to obtain the first analytical estimates of the DM relic abundance in Section \ref{sec:relicdensity}. When further analytical progress is no longer possible, we turn to 3+1 classical lattice simulations, performing a comprehensive scan of the parameter space in Section~\ref{sec:simulations}. Finally, in Section~\ref{sec:discussion} we present our conclusions and discuss possible extensions of the model as well as directions for future research. Throughout this work, we use natural units ($c=\hbar=1$) and adopt the metric signature $(-,+,+,+)$.

\section{\label{sec:model}Gravitational EFT framework} \label{sec:EFT}

We work within an EFT description of gravity coupled to matter. The field content consists of the spacetime metric $g_{\mu\nu}$, an inflaton field $\phi$ responsible for driving the background expansion, and a real scalar field $\chi$, which plays the role of a spectator DM degree of freedom, in the sense that it remains energetically subdominant, induces negligible backreaction on the metric, and has negligible couplings to the SM degrees of freedom. The theory is taken to respect diffeomorphism invariance and a discrete $\mathbb{Z}_2$ symmetry under $\chi\rightarrow -\chi$, ensuring the stability of the dark sector in the absence of additional interactions.

The considered EFT is organized as an expansion in local operators of increasing mass dimension, suppressed by appropriate powers of a scale $\Lambda$ which should be interpreted as the EFT suppression scale and may lie parametrically below the true ultraviolet (UV) cutoff of the theory \cite{Aydemir:2012nz}. At dimension four and below, and restricting to operators relevant for the curvature-induced instability discussed in this work,\footnote{In writing Eq.~\eqref{Srenorm}, we have omitted a quartic self-interaction $\lambda\chi^4$, which is a priori allowed by the $\mathbb{Z}_2$ symmetry. This is a deliberate and  natural simplification. The curvature-induced tachyonic instability that drives particle production is entirely controlled by the quadratic effective mass of the spectator field $\chi$ and is therefore insensitive to small self-interactions at linear level. Moreover, in the absence of direct couplings between the dark and visible sectors, the radiative generation of sizable self-interactions through gravitational effects is parametrically suppressed by powers of $H_*/\Lambda$, with $H_*$ the Hubble scale at the onset of the tachyonic instability, and $\Lambda$ the UV cutoff of the theory. While small self-interactions may become relevant at later stages of the non-linear evolution—potentially affecting the detailed thermalization or fragmentation dynamics—they do not affect the existence, onset, or efficiency of the curvature-driven instability itself and are therefore neglected in this minimal setup.} the spectator sector is described by
\begin{equation}\label{Srenorm}
S_{\leq 4} = \int d^4x\,\sqrt{-g}\,
\bigg[\frac{M_P^2}{2}\,R +\mathcal{L}_\phi -\frac{1}{2}\,g^{\mu\nu}\partial_\mu\chi\,\partial_\nu\chi
-\frac{1}{2}\,M^2\chi^2 -\frac{1}{2}\,\xi\,R\,\chi^2
\bigg]\,,
\end{equation}
with $M_P$ the reduced Planck mass, $M$ the bare mass of $\chi$, and $\xi$ a dimensionless non-minimal coupling to the Ricci scalar. The inflaton sector is assumed to be minimally coupled to gravity and kept general through the Lagrangian density $\mathcal{L}_\phi$, which, for definiteness, may be taken as
\begin{equation}
\mathcal{L}_\phi = -\frac{1}{2}\,g^{\mu\nu}\partial_\mu\phi\,\partial_\nu\phi- V(\phi)\,,
\end{equation}
although our results do not depend on the detailed form of the inflationary potential $V(\phi)$.

At the next order in the EFT expansion, dimension-six operators involving the curvature invariants and the spectator field $\chi$ appear. We restrict ourselves to the leading interactions that: i) respect diffeomorphism invariance and the aforementioned $\mathbb{Z}_2$ symmetry $\chi\to-\chi$, ii) are local, iii) contain at most two derivatives acting on $\chi$, so that the matter sector remains free of higher-derivative instabilities, and iv) allow a transparent separation between the gravitational background dynamics and the dynamics of the spectator field. Under these assumptions, the leading curvature-induced interactions involving $\chi$ take the schematic form $\chi^2 \mathcal{I}$, where $\mathcal{I}$ denotes a scalar constructed from the spacetime curvature. At mass dimension six, a convenient and complete basis for such operators is given by
$\{R^2,\;R_{\mu\nu}R^{\mu\nu},\;R_{\mu\nu\rho\sigma}R^{\mu\nu\rho\sigma}\}$, up to total derivatives and field redefinitions. Higher-curvature invariants are suppressed by additional powers of the cutoff and are neglected within the present truncation. Operators involving additional derivatives acting on the curvature, such as $\chi^2 \Box R$, can either be eliminated by integration by parts and local field redefinitions at the order we work, or correspond to higher-order terms in the derivative expansion. As such, they do not introduce qualitatively new interactions at leading order and will not be considered explicitly in what follows. Similarly, operators with derivatives acting on $\chi$, such as $(\nabla\chi)^2 R/\Lambda^2$ or $G^{\mu\nu}\nabla_\mu\chi\nabla_\nu\chi/\Lambda^2$, modify the kinetic structure of the spectator field. While such terms are a priori allowed by the symmetries of the theory, they can be consistently neglected in a minimal setup focused on curvature-induced interactions of potential type like the one under consideration. Within this EFT truncation, the most general quadratic curvature interaction involving $\chi$ can therefore be written as
\begin{equation} \label{Snonrenorm}
\Delta S =-\int d^4x\,\sqrt{-g}\,
\left(\alpha R^2 +\beta R_{\mu\nu}R^{\mu\nu} +\gamma R_{\mu\nu\rho\sigma}R^{\mu\nu\rho\sigma}\right)\frac{\chi^2 }{\Lambda^2}\,,
\end{equation}
where $\alpha$, $\beta$ and $\gamma$ are dimensionless, real Wilson coefficients. 

\begin{figure}
  \centering
  \includegraphics[width=0.8\textwidth]{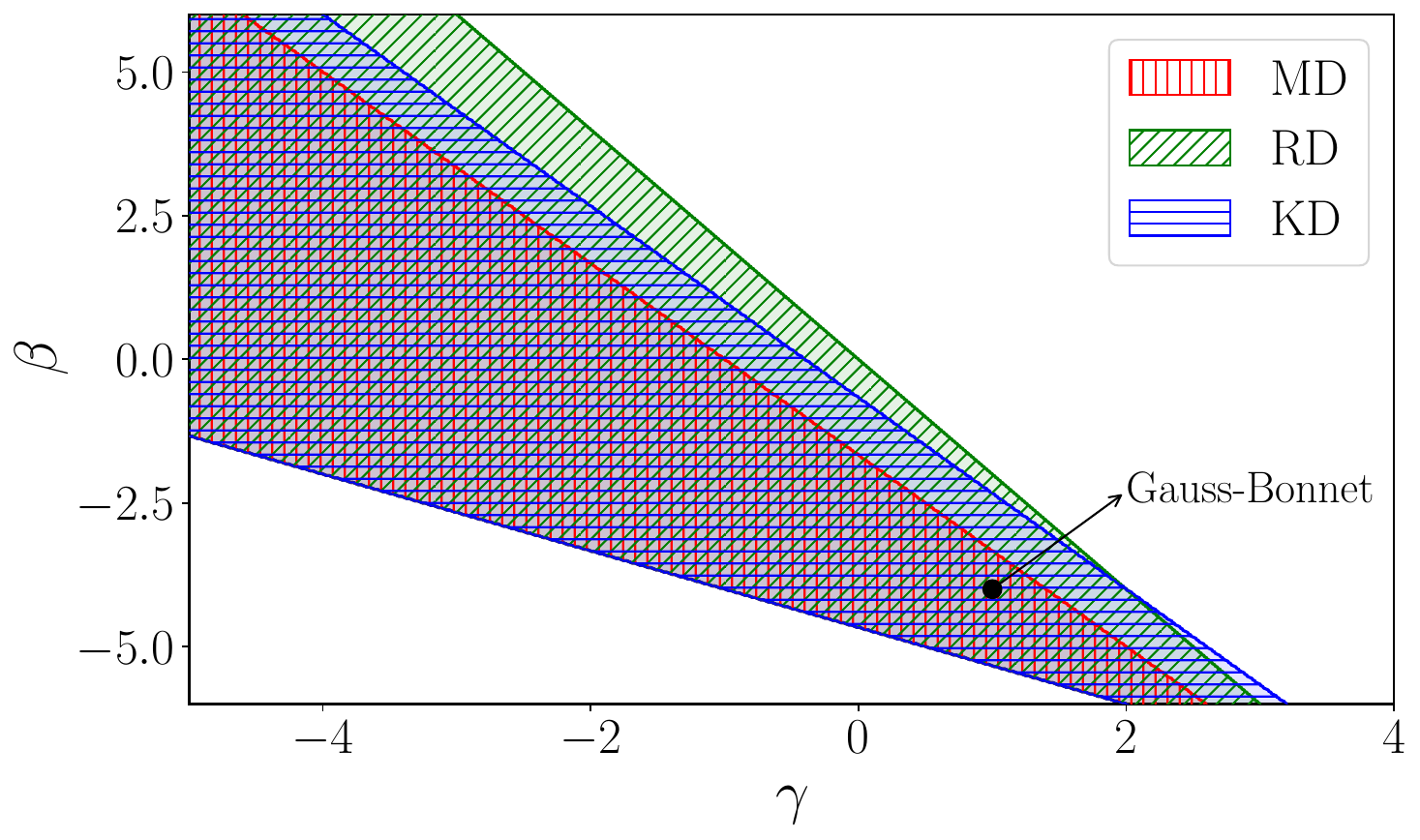}
  \caption{Allowed regions in the $(\gamma,\beta)$ parameter space for which the curvature-induced contribution to $M_{\rm eff}^2$ becomes negative, triggering a spontaneous symmetry breaking of the $\mathbb{Z}_2$ symmetry of the spectator field for sufficiently small bare masses $M$. The regions are shown for fixed $\alpha=1$,  $\xi=4$ and $\Lambda=H_*$,  with $H_*$ the Hubble scale at the onset of the tachyonic instability, and correspond to different post-inflationary cosmological epochs, classified by their average equation of state $w$: Matter Domination (MD), Radiation Domination (RD) and Kinetic Domination (KD). The black dot corresponds to the benchmark Gauss-Bonnet scenario considered in this paper, $(\alpha,\beta,\gamma)=(1,-4,1)$.}
  \label{fig:coefficient_region}
\end{figure}

Collecting Eqs.~\eqref{Srenorm} and \eqref{Snonrenorm}, the dynamics of the spectator field $\chi$ is governed by an effective quadratic action in which the coupling to spacetime curvature induces a background-dependent effective mass
\begin{equation}\label{massgeneric}
M_{\rm eff}^2=M^2+ \xi R+ \frac{2}{\Lambda^2}\left(\alpha R^2+ \beta R_{\mu\nu}R^{\mu\nu}+ \gamma R_{\mu\nu\rho\sigma}R^{\mu\nu\rho\sigma}
\right)\,.
\end{equation}
To make further progress, we specialize to a cosmological background described by a flat Friedmann--Lemaître--Robertson--Walker (FLRW) metric $g_{\mu\nu}=\operatorname{diag}(-1,a^2(t)\delta_{ij})$ with scale factor $a(t)$. In such a homogeneous and isotropic background, all curvature invariants can be expressed algebraically in terms of the Hubble rate $H=\dot a/a$ and the equation-of-state parameter $w$ of the dominant energy component. As a result, the effective mass of the spectator field in \eqref{massgeneric} becomes a time-dependent quantity controlled by the cosmic expansion history, namely 
\begin{equation}
   M_{\rm eff}^2=M^2+ 3\xi(1 - 3w)H^2 + 18\left[\alpha(1-3w)^2 + \beta(1+3w^2) + \frac13\gamma (4 + \left(1+3w\right)^2) \right]\frac{H^4}{\Lambda^2}\,.
    \label{eq:effective_mass}
\end{equation}
This explicit form highlights two important features. First, the curvature-induced contributions are entirely determined by the background expansion and do not rely on direct interactions between the spectator field $\chi$ and other matter fields. Second, different cosmological epochs can lead to qualitatively different behavior for $M_{\rm eff}^2$, depending on the relative importance and signs of the $H^2$ and $H^4$ coefficient (cf. Table \ref{tab:couplings} and Fig. \ref{fig:coefficient_region}). In particular, a spontaneous symmetry breaking will occur if the squared effective mass \eqref{eq:effective_mass} changes from positive to negative values. 

\begin{table}
\centering
\renewcommand{\arraystretch}{1.4}
\begin{tabular}{c c c}
\hline
\textbf{Period} & \textbf{EoS} & \textbf{$M^2_{\rm eff}-M^2$} \\
\hline
Inflation 
& $w=-1$ 
& $12\xi H^2 
+ {18}\!\left(16\alpha+4\beta+\tfrac{8}{3}\gamma\right)
\dfrac{H^4}{\Lambda^2}$ \\[0.3em]
Matter domination 
& $w=0$ 
& $3\xi H^2 
+ {18}\!\left(\alpha+\beta+\tfrac{5}{3}\gamma\right)
\dfrac{H^4}{\Lambda^2}$ \\[0.3em]
Radiation domination 
& $w=\tfrac{1}{3}$ 
& ${24}\left(\beta+2\gamma\right)
\dfrac{H^4}{\Lambda^2}$ \\[0.3em]
Kination 
& $w=1$ 
& $-6\xi H^2 
+{72}\!\left(\alpha+\beta+\tfrac{5}{3}\gamma\right)
\dfrac{H^4}{\Lambda^2}$ \\[0.3em]
\hline
\end{tabular}
\caption{Curvature-induced contribution to the effective mass of the spectator field $\chi$ during different cosmological epochs, evaluated on an FLRW background with  constant equation of state $w$. The expressions show explicitly the $H^2$ and $H^4$ scaling and their dependence on the curvature-coupling coefficients $(\xi,\alpha,\beta,\gamma)$.}
\label{tab:couplings}
\end{table}

From an EFT perspective, the relative importance of the curvature-induced contributions is governed by power counting in $H/\Lambda$, with $H\lesssim \Lambda$. For ${\cal O}(1)$ values of the curvature-coupling coefficients $(\xi,\alpha,\beta,\gamma)$, the contribution to $H^2$ originating from the non-minimal Ricci coupling in Eq.~\eqref{Srenorm} is expected to dominate over the dimension-six term, scaling as $H^4/\Lambda^2$. Consequently, during most cosmological epochs, the sign of $M_{\rm eff}^2-M^2$ is primarily controlled by the factor $(1-3w)$, implying that a sign change between inflation and a subsequent decelerating phase can only occur for sufficiently stiff equations of state, $w>1/3$, associated, for instance, with post-inflationary quintessential inflation epochs \cite{Peebles:1998qn,Spokoiny:1993kt,Wetterich:1987fm,Wetterich:1994bg,Rubio:2017gty,Bernal:2020bfj}. This is the situation giving rise to Hubble-induced phase transitions \cite{Bettoni:2018utf,Bettoni:2018pbl,Bettoni:2019dcw}, Ricci reheating scenarios \cite{Bettoni:2021zhq,Laverda:2023uqv,Laverda:2024qjt,Bettoni:2024ixe,Laverda:2025pmg} (for a review, see e.g. \cite{Bettoni:2021qfs, Rubio:2025egw}) and non-minimal super-horizon modes amplification \cite{Chakraborty:2024rgl, Chakraborty:2025oyj}. In this sense, curvature-induced symmetry breaking driven by quadratic invariants would normally be subleading within a standard EFT hierarchy.

RD constitutes a notable and highly non-generic exception to the previous expectation. Since the equation of state satisfies $w=1/3$, the Ricci scalar vanishes identically and all curvature-induced contributions proportional to $H^2$ drop out of the effective mass \eqref{eq:effective_mass}. As a result, the leading curvature effect is automatically provided by the higher-dimensional term scaling as $H^4/\Lambda^2$, even when the EFT expansion remains well controlled. This makes RD a particularly natural setting in which curvature-squared operators can trigger a transient tachyonic instability without requiring a breakdown of power counting or finely tuned cancellations among operators. It is precisely this property that motivates our focus on symmetry breaking at the transition between inflation and RD and guides the identification of specific curvature combinations capable of realizing this behavior.

In order to have spontaneous symmetry breaking during RD at least one of the Wilson coefficients $\alpha,\beta,\gamma$ in Eq.~\eqref{eq:effective_mass} should be negative. A particularly distinguished case is provided by the Gauss--Bonnet combination, corresponding to $(\alpha,\beta,\gamma)=(1,-4,1)$. For this choice, the quadratic curvature invariants combine into the Gauss--Bonnet density
\begin{equation}
\mathcal{G}
= R^2 - 4 R_{\mu\nu}R^{\mu\nu} + R_{\mu\nu\rho\sigma}R^{\mu\nu\rho\sigma}=\frac{1}{4}\, \epsilon^{\mu\nu\rho\sigma} \epsilon_{\alpha\beta\gamma\delta} R^{\alpha\beta}{}_{\mu\nu} R^{\gamma\delta}{}_{\rho\sigma}\,,
\end{equation}
whose contribution to the effective mass \eqref{eq:effective_mass} in a cosmological FLRW background is proportional to $-(1+3w)H^4$. As a result, the Gauss--Bonnet term yields a positive contribution to $M^2_{\rm eff}$ during inflation ($w\simeq -1$) and a negative contribution during RD ($w=1/3$), namely
\begin{equation} 
M^2_{\rm eff}= 
\begin{cases} 
M^2+12\xi H^2 + 48\frac{H^4}{\Lambda^2}\,, &\quad  \text{Inflation} \\ 
M^2-48 \frac{H^4}{\Lambda^2}\,, &\quad \text{RD} 
\end{cases} \,.
\label{eq:final_action}
\end{equation}
This property ensures that a sign flip in $M_{\rm eff}^2$ at the inflation--radiation transition arises as a robust consequence of the background dynamics, rather than from fine-tuned cancellations among unrelated operators.

An additional motivation for focusing on the Gauss--Bonnet invariant is theoretical consistency. Generic quadratic curvature terms are known to modify the gravitational sector by introducing additional propagating degrees of freedom. In particular, an $R^2$ term leads to an extra scalar mode, while $R_{\mu\nu}R^{\mu\nu}$ generically introduces a massive spin--2 ghost \cite{Sotiriou:2008rp}. Since our aim is not to study modified gravity but rather particle production on a fixed cosmological background, we impose the requirement that the gravitational equations of motion remain second order and that no additional gravitational degrees of freedom are activated within the regime of validity of the EFT.

At this stage, it is important to clarify the regime of validity of the EFT employed throughout this work. The EFT expansion is organized in powers of curvature invariants over the suppression scale $\Lambda$, schematically $R/\Lambda^2 \sim H^2/\Lambda^2$, rather than in powers of the field amplitude $\chi$ itself. 
In particular, the dimension-six operators considered in Eq.~\eqref{Snonrenorm} correct the Hubble-induced effective mass \eqref{eq:final_action} by terms scaling as $H^4/\Lambda^2$, while higher-order curvature operators would enter suppressed by additional powers of $H^2/\Lambda^2$. 
Therefore, as long as no new propagating gravitational degrees of freedom are activated and 
\begin{equation}\label{deltaH}
\delta_{\rm H}\equiv \frac{H^2}{\Lambda^2} \lesssim \mathcal{O}(1)\,,
\end{equation} 
the truncation at dimension six in the scalar-tensor sector of the theory provides a controlled description of curvature-induced effects.~\footnote{Note, however, that the saturation of the EFT ratio $\delta_{\rm H} \sim \mathcal{O}(1)$ does not necessarily imply by itself a breakdown of the effective description. As emphasized e.g. in Ref.~\cite{Aydemir:2012nz}, tree-level unitarity violation and the onset of new degrees of freedom can be parametrically disconnected, and EFTs may self-heal apparent perturbative inconsistencies through loop resummations.} Moreover, since the metric is treated as a fixed FLRW background and the spectator energy density remains subdominant by construction, potentially dangerous higher-curvature operators in the \textit{pure} gravitational sector are never dynamically excited. In fact, for all sub-Planckian field excursions of relevance one finds $M_P^2 R \gg \chi^2\,\mathcal{G}/\Lambda^2$, or equivalently $\delta_{\rm H}\,\chi^2 \ll M_P^2$, ensuring that the Einstein–Hilbert term parametrically dominates over curvature–scalar corrections and that the leading gravitational dynamics remain governed by the usual two-derivative structure.

\section{Tachyonic instability and dark sector evolution} \label{sec:analytics} 

We now turn to the dynamical consequences of the Gauss-Bonnet-induced mass contribution derived in the previous section. In particular, we study the conditions under which the Gauss--Bonnet contribution drives a transient tachyonic instability of the spectator field and analyze the subsequent evolution of the dark sector. The field dynamics follow from the equation of motion obtained by varying the action with respect to the spectator field $\chi$,
\begin{align}
    &\ddot\chi+3H\dot\chi-\frac{1}{a^{2}}\nabla^2\chi+\left(M^2+\frac{2\mathcal{G}}{\Lambda^2}\right)\chi=0\,,
    \label{eq:Field_eq}
\end{align}
with
\begin{equation}
\mathcal{G}=-12(1+3w)H^4\,,
\end{equation} 
the Gauss--Bonnet term for a FLRW metric.  The energy density and pressure associated with the spectator field are obtained from the energy--momentum tensor and given by
\begin{equation}
    \rho_{\chi}=\tensor{T}{_0_0}=\frac{\dot \chi^2}{2}+\frac{(\nabla \chi)^2}{2a^2}+\frac{1}{2}M^2\chi^2-\frac{8H^2}{\Lambda^2}\left(\frac{\nabla^2}{a^2}-3H\partial_t\right)\chi^2\,,
    \label{eq:chi_en}
\end{equation}
\begin{equation}
    p_{\chi}=\frac{1}{3a^2} \sum_i\tensor{T}{_i_i}=\frac{\dot \chi^2}{2}-\frac{(\nabla \chi)^2}{6a^2}-\frac{1}{2}M^2\chi^2-\frac{8H^2}{3\Lambda^2}\left((1+3w)\left(\frac{\nabla^2}{a^2}-3H\partial_t\right)+3\partial_t^2\right) \chi^2\,,
    \label{eq:chi_press}
\end{equation}
with $\partial_t\equiv \partial/\partial t$. The dynamics of the system is characterized by two qualitatively distinct regimes:
\begin{enumerate}
\item \textit{Inflation} ($w=-1$): The effective mass \eqref{eq:final_action} of the spectator field is positive, and $\chi$ is stabilized at the origin of its potential. Provided $M_{\rm eff}\gtrsim \mathcal{O}(5)H$ for the observable numbers of $e$-folds of inflation, isocurvature perturbations are exponentially suppressed \cite{Chung:2004nh}. Assuming $M\ll H$, this condition reduces to a constraint $\xi + 4\left({H}/{\Lambda}\right)^2 \gtrsim 2$ on the curvature--induced contributions. When $H/\Lambda\ll 1$, the dimension--six term is negligible and the bound simply requires $\xi \gtrsim 2$. As $H$ approaches $\Lambda$, higher--dimensional curvature effects provide an additional positive contribution. By contrast, in the minimally coupled limit $\xi\to 0$, achieving the same level of suppression would demand inflationary scales $H \gtrsim \Lambda/\sqrt{2}$, close to the limit of validity of EFT. Overall, for a light spectator field, the condition $M_{\rm eff}\gtrsim \mathcal{O}(5) H$ is fulfilled across a broad and theoretically well--controlled region of parameter space, with modest non--minimal coupling already ensuring exponential damping of isocurvature fluctuations.
\item \textit{Radiation domination} ($w=1/3$): The transition to RD induces a change in the sign of the  effective mass $M_{\rm eff}^2$, triggering a spontaneous breaking of the $\mathbb{Z}_2$ symmetry. The field acquires a tachyonic mass and undergoes a period of rapid growth, driven by the curvature-induced instability. This amplification stage terminates once the curvature contribution becomes subdominant compared to the bare mass term $M^2$, after which the symmetry is restored as the Universe expands.
\end{enumerate}
\noindent We will assume that the shift from inflation to a decelerating expansion occurs abruptly, as happens for instance in some T-models of inflation \cite{Lozanov:2019ylm,Rubio:2019ypq,Dux:2022kuk}. This is a valid approximation if the inflaton decays into radiation nearly instantaneously at some time $t_*$, with a decay timescale much shorter than the Hubble time at that moment ($H^{-1}$). Under such circumstances, treating the transition as instantaneous is justified.~\footnote{Reheating can in general be a prolonged and model-dependent process, in which case the background evolution may deviate from the idealized instantaneous transition adopted here; for a review see e.g. \cite{Barman:2025lvk}. Many studies of gravitational DM production therefore focus on specific inflationary realizations, including chaotic inflation with quadratic
\cite{Markkanen:2015xuw,Fairbairn:2018bsw,Cembranos:2019qlm,Kainulainen:2022lzp,Kainulainen:2024etd,Laulumaa:2020pqi,Lebedev:2022vwf} or quartic
\cite{Fairbairn:2018bsw,Babichev:2020yeo,Lebedev:2022vwf} potentials, Starobinsky-like models \cite{Ema:2016hlw,Laulumaa:2020pqi,Verner:2024agh}, hilltop scenarios \cite{Ema:2018ucl}, as well as hybrid and natural inflation frameworks \cite{Chung:2001cb}. In such settings, the reheating or preheating phase may feature rapid oscillations of the inflaton, which become imprinted onto curvature quantities such as the Ricci scalar and the Hubble rate. This can induce additional resonant effects in the spectator sector. However, these are typically of the parametric type and generically much less efficient than the tachyonic instabilities considered in this work. A faithful treatment of such effects requires a fully specified inflationary model and lies beyond the scope of the present analysis.} 

To proceed analytically, we adopt a simple parametrization of the background expansion that captures the essential features of a Universe dominated immediately after inflation by a radiation fluid with a constant equation of state $w=1/3$, namely
\begin{equation}
    a(t) =a_*\left[1+2H_*(t-t_*)\right]^{1/2},
    \qquad\quad 
    H(t)=\frac{\dot a}{a}= \frac{H_*}{1+2H_*(t-t_*)}\,,
    \label{eq:param_H_a}
\end{equation}
with $a_*$ and $H_*$ denoting the scale factor and Hubble rate evaluated at a reference time $t_*$ marking the onset of the RD era. In order to simplify the subsequent analysis, it is also convenient to work with dimensionless variables that absorb the usual effects of the background expansion and isolate the physical instability of the spectator field. We therefore introduce the rescaled field and coordinates
\begin{equation}
    Y \equiv \frac{a}{a_*}\frac{\chi}{\chi_*}\,,
    \qquad   \qquad
    \vec y \equiv a_*\chi_*\,\vec x\,,
    \qquad   \qquad
    z \equiv a_*\chi_*\,\tau\,,
\end{equation}
where $\tau=\int dt/a$ is the conformal time and 
\begin{equation}\label{chistardef}
\chi_*=\sqrt{48}\,H_*^2/\Lambda
\end{equation}
defines the characteristic mass scale associated with the curvature-induced instability. In terms of these variables, the background evolution takes a particularly simple form. The scale factor and the conformal Hubble rate can be expressed as
\begin{equation}
    a(z)=a_*\left(1+\frac{z}{\nu}\right)\,,
    \qquad    \qquad
    \mathcal{H}(z)=   \frac{a'}{a}=    \frac{1}{z+\nu}\,,
    \qquad    \qquad
    \nu    = \frac{\chi_*}{H_*}\,,
\end{equation}
with primes denoting derivatives with respect to $z$. Choosing the origin of conformal time such that $z(t_*)=0$, the equation of motion for the rescaled field $Y$ reduces to a Klein--Gordon equation 
\begin{equation}
    Y'' - \nabla^2 Y - M_{\rm eff}^2(z)\,Y = 0\,,
    \label{eq:KG_Y}
\end{equation}
with time-dependent effective mass
\begin{equation}
    M_{\rm eff}^2(z) =
    \left(\frac{1}{1+z/\nu}\right)^6 -\frac{M^2}{\chi_*^2}\left(1+\frac{z}{\nu}\right)^2\,.
    \label{eq:mass}
\end{equation}
In the following sections, we analyze Eq.~\eqref{eq:KG_Y}, beginning with a homogeneous classical approximation before turning to the study of quantum fluctuations.

\subsection{Classical homogeneous dynamics} \label{sec:homo} 

A first qualitative understanding of the system can be obtained by treating the DM spectator field as a classical, fully coherent degree of freedom, therefore neglecting its true quantum nature. Under this assumption, spatial gradients can be ignored and the dynamics is governed by the temporal evolution of a homogeneous condensate. The field equation \eqref{eq:KG_Y} then reduces to
\begin{equation}
    Y'' - M_{\rm eff}^2(z)\,Y = 0\,,
    \label{eq:KG_Y_NoGradient}
\end{equation}
with initial conditions $Y(0)\ll 1$ and $Y'(0)=2Y(0)/\nu$ following, for $\xi \gg 1$, from $\chi(t_*)\sim {H_*}/{(2\pi\sqrt{12\xi})}$ and $\dot{\chi}(t_*)\simeq H_*\,\chi(t_*)$ \cite{Bettoni:2018utf}.
\begin{figure}
    \centering
    \includegraphics[width=0.9\textwidth]{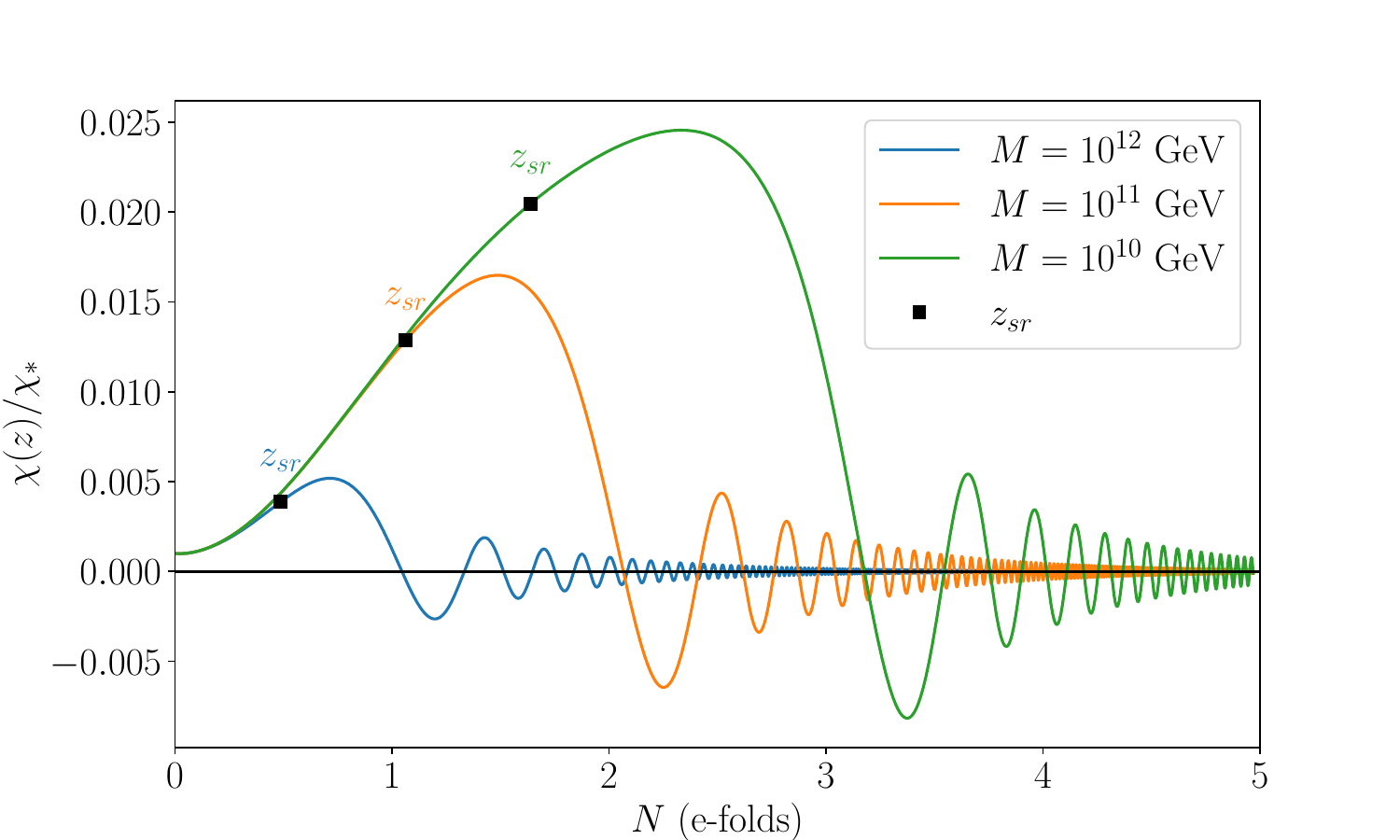}
    \caption{Time evolution of the homogeneous field amplitude, normalized to $\chi_*$ (cf. \eqref{chistardef}), as a function of the number of $e$-folds since the onset of RD, $N\equiv\ln(a/a_*)=\ln(1+z/\nu)$ for different values of the bare mass $M$, obtained from numerically solving Eq.~\eqref{eq:KG_Y_NoGradient}. The parameters are fixed to $H_*=\Lambda=10^{12}\,\mathrm{GeV}$ with initial conditions $Y(0)=10^{-3}$ and $Y'(0)=2Y(0)/\nu$, following from the inflationary expectation $\chi(t_*)\simeq {H_*}/{(2\pi\sqrt{12\xi})}$ and $\dot{\chi}(t_*)\simeq H_*\,\chi(t_*)$, with $\xi\simeq 45$ \cite{Bettoni:2018utf}. Black squares indicate the symmetry restoration time $z_{\rm sr}$ at which the effective squared mass \eqref{eq:mass} changes sign.}
    \label{fig:homogeneous_sol}
\end{figure}
\begin{figure}
    \centering
    \includegraphics[width=0.9\textwidth]{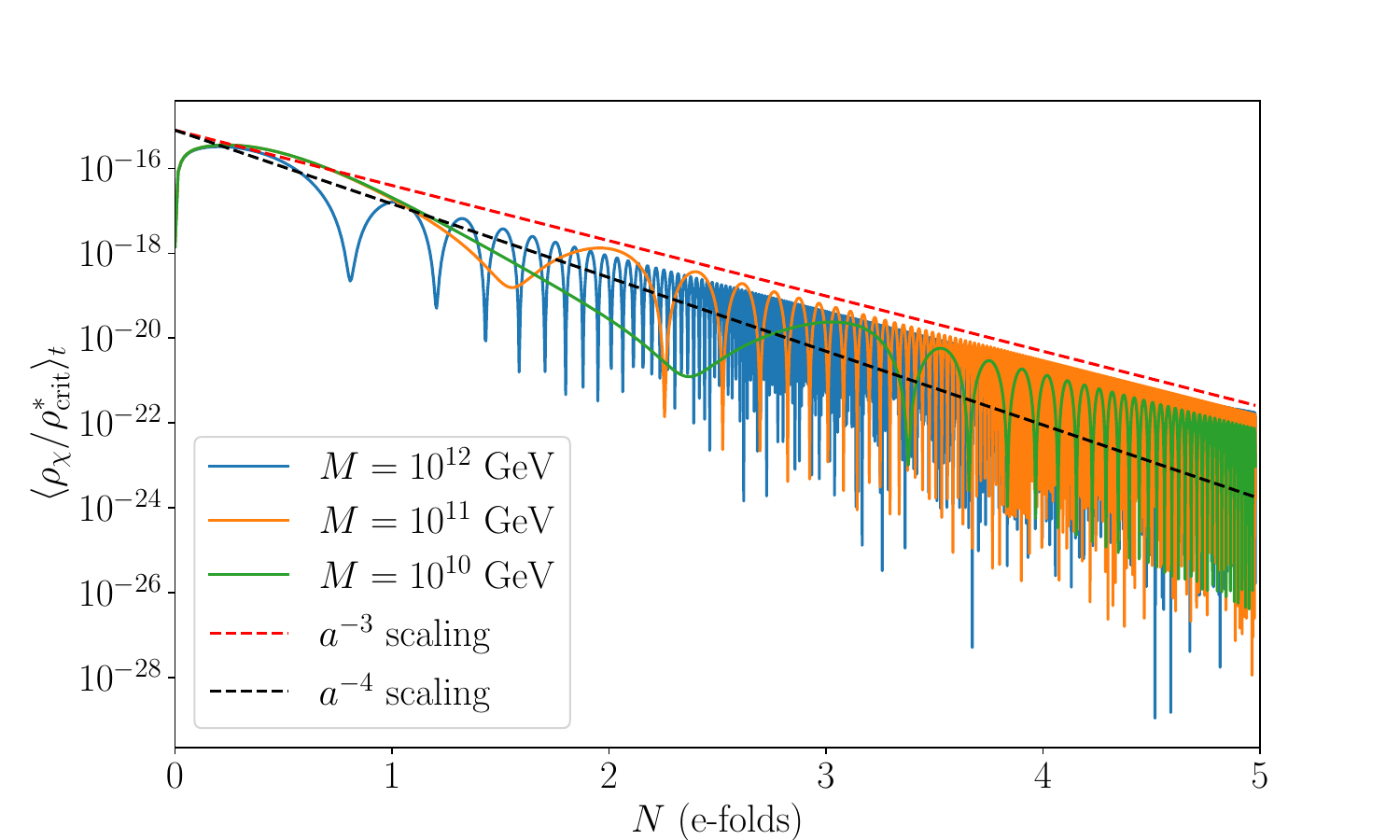}
    \caption{Evolution of the homogeneous energy density of the DM field as a function of the number of $e$-folds since the onset of RD, $N\equiv\ln(a/a_*)=\ln(1+z/\nu)$, for different values of the bare mass $M$, computed from Eq.~\eqref{eq:chi_en} and normalized to the critical density $\rho_{\rm crit}=3M_P^2H_\ast^2$. The parameters are fixed to $H_*=\Lambda=10^{12}\,\mathrm{GeV}$, with initial conditions $Y(0)=10^{-3}$ and $Y'(0)=2Y(0)/\nu$ motivated by inflationary expectations for the spectator field, namely  $\chi(t_*)\simeq {H_*}/{(2\pi\sqrt{12\xi})}$ and $\dot{\chi}(t_*)\simeq H_*\,\chi(t_*)$, with $\xi\simeq 45$ \cite{Bettoni:2018utf}. The red and black dashed lines indicate, respectively, the reference scalings $\rho\propto a^{-3}$ characteristic of pressureless matter and $\rho\propto a^{-4}$ characteristic of radiation}
    \label{fig:homogeneous_endens}
\end{figure}
Figure~\ref{fig:homogeneous_sol} shows numerical solutions of this equation for several values of the bare mass parameter $M$, displayed as a function of the number of $e$-folds $N\equiv\ln(a/a_*)=\ln(1+z/\nu)$ since the onset of RD. Note that in this and the following illustrative plots of this section we set $\Lambda = H_*$ in order to display the maximal deviation from the minimal scenario. This choice should be regarded, however, as a boundary benchmark rather than a fully controlled EFT regime; all quantitative conclusions in Section \ref{sec:simulations} are instead drawn from $\Lambda \gtrsim \mathrm{few}\,H_*$, where $\delta_{\rm H} \ll 1$.

During the early stages, the effective mass squared is negative and the field undergoes a rapid tachyonic amplification. As the Universe expands, the curvature-induced contribution to $M_{\rm eff}^2$ is progressively diluted, eventually restoring a positive effective mass. In the homogeneous approximation, this moment corresponds to the inflection point of the field evolution and defines a characteristic timescale
\begin{equation}
    z_{\rm sr}=\nu\!\left[\left(\frac{\chi_*}{M}\right)^{1/4}-1\right]\,,
    \label{eq:z_sr}
\end{equation}
following directly from Eq.~\eqref{eq:mass}. As expected, lighter fields experience a longer amplification phase, since the curvature-induced contribution takes more time to redshift below the bare mass term $M^2$.

After symmetry restoration, the field undergoes coherent oscillations around the minimum of its quadratic potential. In the homogeneous limit, a scalar field oscillating in a potential $V(\phi)\propto\phi^{2n}$ is characterized by an averaged equation of state $\langle w\rangle=(n-1)/(n+1)$ \cite{Turner:1983he}. Although this relation is only approximate in the present setup—since the definitions of energy density and pressure in Eqs.~\eqref{eq:chi_en} and \eqref{eq:chi_press} involve spatial derivatives—it provides a useful guideline for interpreting the late-time behavior of the system \cite{Lozanov:2016hid}. The evolution of the energy density, shown in Fig.~\ref{fig:homogeneous_endens} for different bare masses, confirms this expectation: at late times the energy density asymptotically scales as $a^{-3}$, while the pressure rapidly approaches zero. The homogeneous condensate behaves therefore as an effectively pressureless matter component after the tachyonic phase, making it a priori a viable DM candidate.

\subsection{Quantum fluctuations and mode evolution }\label{sec:quantum}

The homogeneous analysis of the previous subsection provides a useful first insight into the instability, but it necessarily neglects the role of unavoidable quantum fluctuations. During a tachyonic phase, these fluctuations are expected to grow rapidly and to dominate the subsequent evolution, rendering the dynamics intrinsically non-homogeneous.

In this subsection we quantize the spectator field $Y$ and study the evolution of its modes in the tachyonic regime. Without attempting to obtain an exact solution of the full dynamics, we show that the amplification of infrared (IR) modes provides the dominant contribution to the field variance after the symmetry breaking, while the homogeneous component plays a subleading role. Our treatment closely follows the standard analyses of tachyonic instabilities in expanding backgrounds \cite{Polarski:1995jg,Lesgourgues:1996jc,Kiefer:1998qe,Garcia-Bellido:2002fsq,Bettoni:2019dcw}. In particular, we treat the metric as classical while quantizing the spectator field $Y$ in Eq.~\eqref{eq:mass}. This approximation is justified since the energy density stored in $\chi$ remains subdominant during the epochs of interest and does not affect the background expansion.

Canonical quantization is implemented by promoting the field $Y$ in Eq.~\eqref{eq:mass}  to a quantum operator $\hat Y$ defined on a fixed classical background and imposing equal-time commutation relations with its conjugate momentum $\hat\Pi=\hat Y'$, namely 
\begin{align}
[\hat Y_{\vec{x}}(z),\hat \Pi_{\vec{y}}(z)] = i\delta^3(\vec{x}-\vec{y})\,, &&
[\hat Y_{\vec{x}}(z),\hat Y_{\vec{y}}(z)] = [\hat \Pi_{\vec{x}}(z),\hat \Pi_{\vec{y}}(z)]=0\,. 
\end{align}
Exploiting spatial translational invariance, the field operator and its conjugate momentum can be expanded in Fourier modes, 
\begin{equation}
    \hat{Y}_{\vec{\kappa}}(z)=f_\kappa(z) \hat{a}_{\vec{\kappa}}(z_0)+f^*_\kappa(z) \hat{a}^{\dagger}_{-\vec{\kappa}}(z_0)~\,,\quad 
    \hat{\Pi}_{\vec{\kappa}}(z)=-i\left(g_\kappa(z) \hat{a}_{\vec{\kappa}}(z_0)-g^*_\kappa(z) \hat{a}^\dagger_{-\vec{\kappa}}(z_0)\right)\,, 
    \label{eq:field_expansion}
\end{equation}
with 
\begin{equation}
\kappa\equiv\frac{|\vec{k}|}{a_*\chi_*}
\end{equation} 
a dimensionless comoving wavenumber and $\hat{a}^\dagger$ and $\hat{a}$ initial time creation and annihilation operators satisfying the canonical commutation relations $
[\hat a_{\vec{\kappa}},\hat a^{\dagger}_{\vec{\kappa}'}] = (2\pi)^3\delta^3(\vec{\kappa}-\vec{\kappa}')
$.  
Since we are dealing with a purely quadratic theory, the Fourier modes $f_\kappa(z)$ decouple and satisfy a harmonic oscillator equation with time-dependent
frequency,
\begin{equation} \label{eq:quantum_freq}
   f_\kappa'' + \omega_\kappa^2(z)\,f_\kappa = 0\,, \qquad\quad
   \omega_\kappa^2(z) = \kappa^2 - M_{\rm eff}^2(z)\,.
\end{equation}
Whenever $\omega_\kappa^2(z)<0$, the corresponding mode $\kappa$ undergoes tachyonic amplification. Focusing on the tachyonic phase $z<z_{\rm sr}$, during which the bare mass is negligible compared to the curvature-induced contribution and approximating
\begin{equation}\label{Mapprox}
  M_{\rm eff}^2(z) \simeq   M^2(z) \equiv  \left(\frac{1}{1+z/\nu}\right)^6\,, \qquad z<z_{\rm sr}\,,
\end{equation}
the condition  $\omega_\kappa^2(z)<0$ defines a time-dependent instability band in momentum space, namely
\begin{equation}  \label{eq:momentum_interval}
        \kappa_{\rm min}(z) < \kappa < \kappa_{\rm max}(z)\,, \qquad\quad
        \kappa_{\rm min}(z)=\mathcal{H}(z)\,, \qquad\quad
        \kappa_{\rm max}(z)=\nu^3\mathcal{H}^3(z)\,,
\end{equation}
with modes with $\kappa>\kappa_{\rm max}(0)$ never experiencing amplification. As the Universe expands and the curvature-induced mass term redshifts away, modes progressively exit the instability band and their growth terminates. The tachyonic amplification is therefore both scale-dependent and transient. 

\begin{figure}
  \centering
    \includegraphics[width=0.9\textwidth]{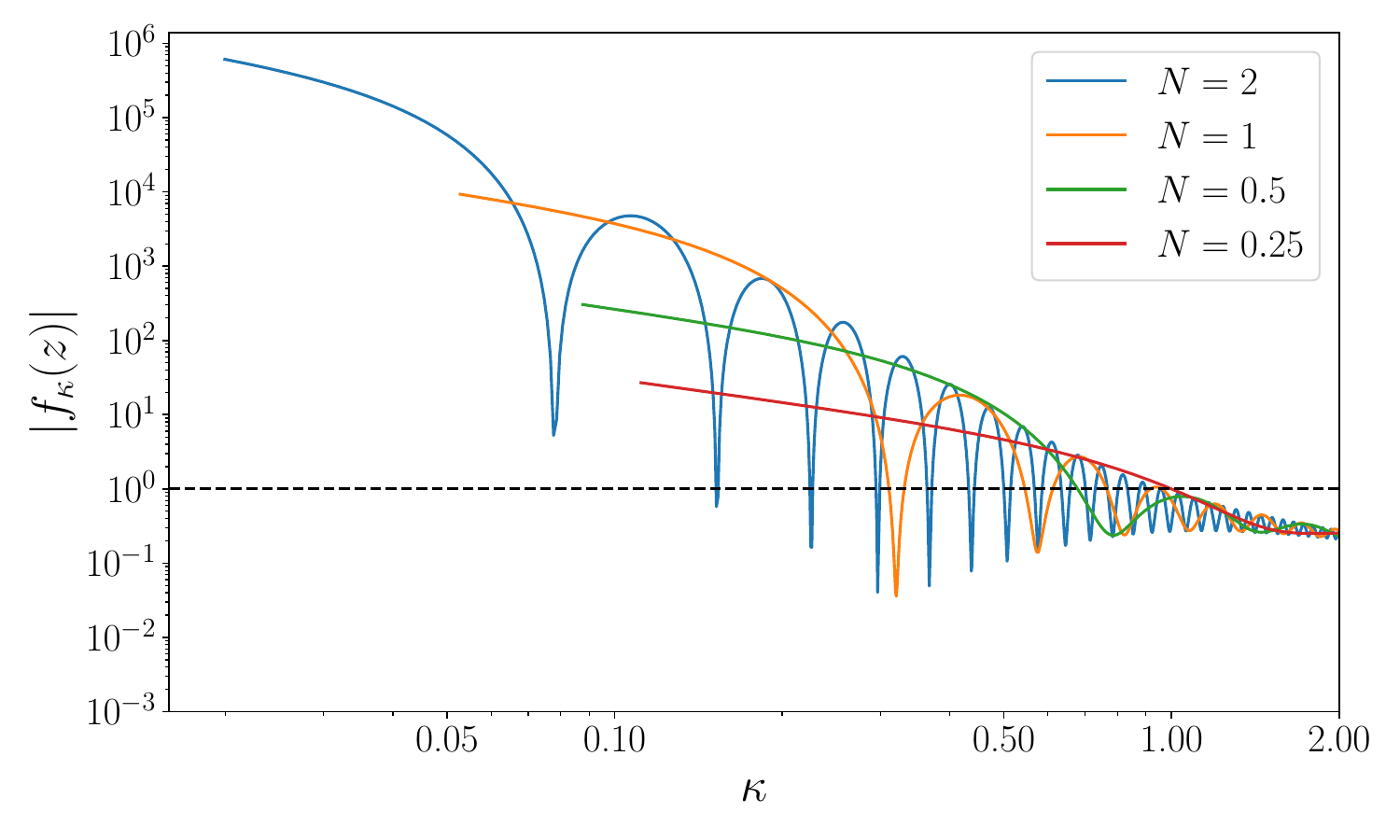}
  \caption{Evolution of the mode amplitudes $|f_\kappa(z)|$  at $N=0.25, 0.5 ,1,$ and $2$ $e$-folds since the onset of RD, for $\Lambda=H_*$. The black horizontal line marks unit normalization. IR modes at low $\kappa$ exhibit pronounced amplification, reflecting the tachyonic growth induced by the curvature-driven instability.}
    \label{fig:f_modes}
\end{figure}

A numerical solution of the mode equation \eqref{eq:quantum_freq}, with vacuum initial conditions $f_\kappa(0)=1/\sqrt{2\kappa}$ and $f_\kappa'(0)=-i\sqrt{\kappa/2}$ imposed at the end of inflation, is displayed in Fig.~\ref{fig:f_modes} for several stages of the subsequent evolution. The results clearly exhibit a well-defined IR amplification at $\kappa < \kappa_{\rm max}(z)$, where modes grow exponentially as a result of the tachyonic instability. Modes in the intermediate range $\kappa_{\rm max}(z) < \kappa < \kappa_{\rm max}(0)$ correspond to fluctuations that were amplified at earlier times but have since then exited the instability band and undergo now oscillatory evolution. Finally, UV modes with $\kappa > \kappa_{\rm max}(0)$ remain unaffected throughout the entire evolution.

The qualitative picture provided by the instability band in Eq.~\eqref{eq:momentum_interval} can be sharpened by constructing an approximate analytical solution for the mode functions. This is particularly useful for understanding the shape and characteristic scales of the amplified power spectrum $\Delta_{\chi}\equiv\kappa^3|f_\kappa|^2$.  The key observation is that the mode equation~\eqref{eq:quantum_freq} admits an approximate turning point $\bar z(\kappa)\simeq\nu(\kappa^{-1/3}-1)$ defined by $\omega_\kappa(\bar z)=0$, which separates an oscillatory regime from a tachyonic one. In the vicinity of this turning point, the adiabaticity condition required for the validity of the WKB approximation necessarily breaks down, and the standard oscillatory and tachyonic WKB solutions cannot be matched reliably. This issue is generic for systems undergoing a transient instability and is independent of the detailed functional form of $M^2(z)$. A consistent way to overcome this limitation is provided by the so-called \textit{uniform approximation} across a turning point, originally developed in Ref.~\cite{OLVER1974392} and applied to cosmological perturbations in Ref.~\cite{Habib_2002}. Sufficiently close to a simple zero of $\omega_\kappa^2(z)$, any second-order linear differential equation of the form $f''+\omega^2(z)f=0$ can be mapped into an Airy differential equation by a suitable change of variables and a rescaling of the dependent function. As a result, Airy functions provide a canonical basis that smoothly interpolates between oscillatory behavior on one side of the turning point and exponential growth on the other. Guided by this observation, we seek a solution of the form
\begin{equation}\label{ansatz}
f_\kappa(z)=\frac{  \mathcal{A}(z)}{\sqrt{2\kappa}}\,W(\mu(z))\,,
\end{equation}
where $W$ is chosen to satisfy the Airy equation
\begin{equation}\label{canonAiry}
\frac{d^2W}{d\mu^2}-\mu W=0\,,
\end{equation}
with independent solutions $\mathrm{Ai}(\mu)$ and $\mathrm{Bi}(\mu)$. Substituting the ansatz \eqref{ansatz} into the mode equation \eqref{eq:quantum_freq} and denoting derivatives with respect to $z$ by primes and derivatives with respect to $\mu$ by subscripts, we obtain
\begin{equation}\label{modeeqW}
(\mu')^2 W_{\mu\mu}
+\left(2\frac{  \mathcal{A}'}{  \mathcal{A}}\mu'+\mu''\right)W_\mu
+\left(\frac{  \mathcal{A}''}{  \mathcal{A}}+\omega_\kappa^2\right)W=0\,.
\end{equation}
The structure of this equation makes it clear that the presence of the first-derivative $W_\mu$ prevents a direct identification of $W$ with the Airy equation. The uniform approximation proceeds by choosing the functions $  \mathcal{A}(z)$ and $\mu(z)$ such that that term vanishes identically. Imposing $2\,  \mathcal{A}'/  \mathcal{A} \,\mu'+\mu''=0$, one finds $ \mathcal{A}^2\mu'=C$, where the integration constant $C$ can be set to unity without loss of generality, as it can be always absorbed consistently into the definition of $\mu$ and the other integration constants appearing in the solution. With this particular choice, the mode equation \eqref{modeeqW} simplifies to
\begin{equation}
(\mu')^2 W_{\mu\mu}+\left(\frac{  \mathcal{A}''}{\mathcal{A}}+\omega_\kappa^2\right)W=0\,. 
\end{equation}
The term $  \mathcal{A}''/  \mathcal{A}$ in this expression represents a subleading adiabatic correction within the uniform approximation and can be therefore consistently neglected at leading order.  Requiring the remaining expression to reduce to the canonical Airy form \eqref{canonAiry}, fixes the remaining functional dependence through the relation $\omega_\kappa^2\simeq(\mu')^2\mu$, which combined with the condition $  \mathcal{A}^2\mu'=1$ and the requirement $\mu(\bar z)=0$, gives
\begin{equation}
  \mathcal{A}(z)=\left(\frac{\mu(z)}{-\omega_\kappa^2(z)}\right)^{1/4}\,, \quad \quad 
\mu(z)=\mathrm{sign}(\bar z-z)
\left[\frac{3}{2}\int_z^{\bar z} ds\,\sqrt{|\omega_\kappa^2(s)|}\right]^{2/3}\,.
\end{equation} 
The resulting uniform approximation for the mode function $f_\kappa(z)$ becomes therefore
\begin{equation}\label{uniformapp}
f_\kappa(z)\simeq\frac{\mathcal A(z)}{\sqrt{2\kappa}}
\left[C_1\,\mathrm{Ai}(\mu(z))+C_2\,\mathrm{Bi}(\mu(z))\right]\,.
\end{equation}
The coefficients $C_1$ and $C_2$ are fixed by imposing vacuum initial conditions $f_\kappa(0)=1/\sqrt{2\kappa}$ and $f_\kappa'(0)=-i\sqrt{{\kappa}/{2}}$  at the onset of RD.  Using the previously derived identity $  \mathcal{A}^2\mu'=1$, the time derivative of the mode function can be written solely in terms of $  \mathcal{A}$, $\mu$, and derivatives with respect to $\mu$. Solving the resulting linear system of equations and making use of the Airy Wronskian $\mathrm{Ai}(\mu)\mathrm{Bi}'(\mu)-\mathrm{Ai}'(\mu)\mathrm{Bi}(\mu)={\pi^{-1}}$, we find
\begin{equation}
C_1=\frac{\pi}{  \mathcal{A}_0}\,\mathcal L_0\,\mathrm{Bi}(\mu_0)\,,
\qquad\qquad
C_2=-\frac{\pi}{  \mathcal{A}_0}\,\mathcal L_0\,\mathrm{Ai}(\mu_0)\,,
\end{equation}
with  $\mathcal L_0\equiv\partial/\partial \mu+\mathcal A_0(\mathcal A'_0+i\kappa \mathcal A_0)$ and the subscript $0$ referring to quantities evaluated at that initial time.

In the tachyonic regime, the solution \eqref{uniformapp} is dominated by the $C_1$ contribution, which is itself controlled by the $\mathrm{Bi}'(\mu_0)$ term. As a result, the $\mathrm{Bi}(\mu(z))$ branch remains subleading and can be consistently neglected, allowing us to set $C_2\simeq 0$. We therefore obtain
\begin{equation}
|f_\kappa|^2\approx \frac{\pi^2}{2\kappa}
        \frac{\omega_{\kappa}(0)}{\omega_{\kappa}(z)} \left( \frac{\mu(z)}{\mu_0}\right)^{1/2}  \mathrm{Bi}'(\mu_0)^2  \mathrm{Ai}(\mu(z))^2     \approx \frac{1}{\kappa} e^{\frac{4}{3}\mu_0^{3/2}} 
    \mathcal P(\kappa,z)\,,
    \label{eq:spectrum_UA}
\end{equation} 
where in the last step we have taken into account the asymptotic expression for the derivative of the Airy function of second kind $\mathrm{Bi}'(\mu_0)\simeq \mu_0^{1/4}e^{\frac{2}{3}\mu_0^{3/2}}/\sqrt{\pi}$ with $\mu_0^{3/2}$ the incomplete Beta-function integral
\begin{align}\label{betafunction}
\mu_0^{3/2}=\frac32 \int_0^{\bar z(\kappa)} \, ds \,\sqrt{M^2(s)-\kappa^2}
=\frac{\nu}{2} \kappa^{2/3}\int_\kappa^{1}  dt\,t^{-5/3}\sqrt{1-t^2}\,,
\end{align}
and defined  
\begin{equation}
t\equiv \kappa (1+s/\nu)^3 \,,\qquad \qquad \mathcal P(\kappa,z)\equiv  \frac{\pi}{2} \frac{\omega_\kappa(0)}{\omega_\kappa(z)}
\left[\mu^{1/4}(z)\mathrm{Ai}(\mu(z))\right]^2\,.
\end{equation}
Equation~\eqref{eq:spectrum_UA} makes manifest the separation between the exponential enhancement associated with the tachyonic phase and the residual prefactors controlling the detailed time and momentum dependence of the spectrum. The dominant contribution arises from the factor $\exp({4}/{3}\,\mu_0^{3/2})$, which encodes the integrated effect of the negative-frequency region experienced by each mode prior to the turning point. This exponential term controls the overall amplification of the spectrum and is responsible for the strong IR enhancement characteristic of the tachyonic instability. By contrast, the remaining factors in ${\cal P}(\kappa,z)$ vary only algebraically with time and momentum and therefore play a subdominant role in shaping the spectrum.

With the above hierarchy in mind, it is instructive to examine the limiting behavior of $\mu_0(\kappa)$ in the far IR and near the edge of the instability band. For sufficiently small momenta, $\kappa\ll1$, the integrand in Eq.~\eqref{betafunction} is dominated by the power-law contribution $t^{-5/3}$, since $\sqrt{1-t^2}\simeq 1$ over most of the integration range. In this regime. the integral can be evaluated analytically, yielding
\begin{equation} \label{mu0smallk}
\mu^{3/2}_0(\kappa)\simeq \frac{3}{4}\,\nu\,(1-\kappa^{2/3})\,,\qquad\quad \kappa\ll 1\,.
\end{equation}
This result shows explicitly that $\mu_0$ approaches a constant of order $\nu$ in the deep IR, leading to an essentially $\kappa$-independent exponential enhancement of long-wavelength fluctuations. By contrast, for modes close to the boundary of the instability band, the square-root factor suppresses the integrand near the upper endpoint, $\sqrt{1-t^2}\simeq\sqrt{2(1-t)}$, while the remaining power-law dependence becomes subleading and can be approximated as $t^{-5/3}\simeq1$. The integral then reduces to
\begin{equation}\label{mu0kone} 
\mu^{3/2}_0(\kappa)\simeq \frac{\sqrt{2}}{3}\,\nu\,(1-\kappa)^{3/2}\,, \qquad \quad \kappa\to 1^-\,.
\end{equation}
In this regime, the exponential enhancement rapidly shuts off as $\kappa$ approaches unity, reflecting the fact that modes near the edge of the instability band experience only a short-lived tachyonic phase.

\begin{figure}
\centering
    \includegraphics[width=0.8\textwidth]{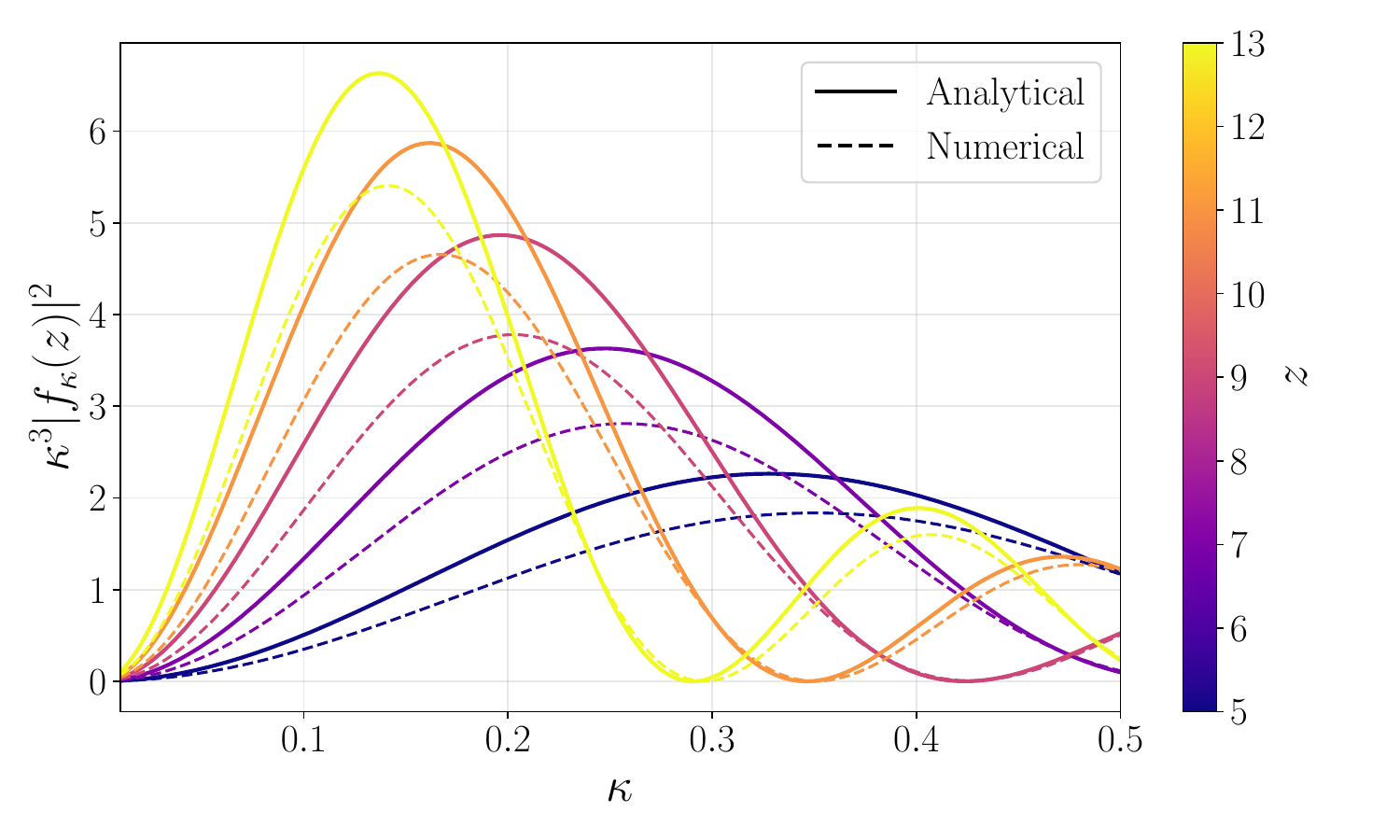}
  \caption{Evolution of the dimensionless power spectrum $\kappa^{3}\lvert f_\kappa(z)\rvert^{2}$ during the tachyonic phase for several times $z=5$--$13$, $\Lambda=H_*$ and $M=10^9$ GeV. Solid curves correspond to the analytical approximation \eqref{spectrumapp} based on the uniform WKB treatment, while dashed curves show the full numerical solution of the mode equation \eqref{eq:quantum_freq}. The spectrum is dominated by IR modes and exhibits a pronounced peak at $\kappa\simeq\kappa_*$, whose amplitude grows rapidly with time.}
  \label{fig:powspectrum}
\end{figure}

The comparison between the power spectrum 
\begin{equation}\label{spectrumapp}
\Delta_{\chi}\equiv\kappa^3|f_\kappa|^2  \approx \kappa^2 e^{\frac{4}{3}\mu_0^{3/2}} \mathcal P(\kappa,z)\,,
\end{equation}
following from the analytical expression \eqref{eq:spectrum_UA} and the one resulting from numerically solving the mode equation \eqref{eq:quantum_freq} is shown in Fig.~\ref{fig:powspectrum} for different times $z$. The uniform approximation accurately captures the location and time dependence of the momentum scale $\kappa_*$ at which the spectrum is maximally amplified, which is the physically relevant quantity for the present analysis. Indeed, since the tachyonic instability is strongly IR dominated, the subsequent dynamics is controlled primarily by modes near $\kappa_*$ rather than by the detailed shape of the spectrum at larger momenta.~\footnote{Correspondingly, the integral of the numerical power spectrum over the amplified region $0<\kappa<1$ is reproduced by the uniform approximation to within a $10\%$ accuracy.} The precise value of $\kappa_*$ is determined by the competition between phase-space suppression effects at small momenta and the exponential tachyonic growth. To make this more explicit, it is useful to introduce the logarithmic spectrum
\begin{equation}\label{kappastareq}
\mathcal S(\kappa)\equiv \ln\Delta_{\chi} = 2\ln\kappa+\frac{4}{3}\,\mu_0^{3/2}(\kappa)+ \ln {\cal P}(\kappa,z)\,.
\end{equation}
Since $\ln{\cal P}$ varies only algebraically with $\kappa$ across the instability band, while the tachyonic enhancement is controlled by the exponential term and scales as $\mu_0^{3/2}(\kappa)=\nu\,q(\kappa)$ with $q(\kappa)$ a dimensionless function, we can determine the location of the peak to leading exponential accuracy by neglecting $\partial_\kappa\ln{\cal P}$. The saddle-point condition $d\mathcal S/d\kappa=0$ then yields
\begin{equation}
\frac{2}{\kappa_*}
+\frac{4}{3}\,\nu\,\frac{dq(\kappa_*)}{d\kappa}\simeq 0\,.
\label{saddleapprox}
\end{equation}
Within the instability band $\kappa\in(0,1)$, the function $q(\kappa)$ is a smooth, order-one function, with no large enhancements or suppressions away from the boundaries. For $\nu=\mathcal O(1)$, corresponding to $\Lambda\sim H_*$, the saddle-point equation \eqref{saddleapprox} admits an interior solution with $\kappa_*=\mathcal O(10^{-1})$ or smaller, depending on the detailed shape of $q(\kappa)$ and the dropped $\cal P$ factor. By contrast, for $\nu\ll1$, the tachyonic term in Eq.~\eqref{saddleapprox} is parametrically suppressed and it cannot generically compensate the phase-space contribution $2/\kappa_*$ within the bulk of the band. Consequently, particle production is strongly suppressed when $\nu\ll1$. 

This dynamical selection of an interior momentum scale $\kappa_*$ has an important implication for the consistency of the EFT description. 
Although the instability band formally extends up to $\kappa_{\rm max}(0)=1$, 
the energy density is dominated by modes around the peak $\kappa_* \ll 1$ determined by Eq.~(3.33).  The physically relevant momentum scale is therefore $p_* \sim \kappa_* \chi_*$,  rather than the maximal scale $\chi_*$ appearing in the definition of the dimensionless variables. Since $\kappa_* = \mathcal{O}(10^{-2}\text{--}10^{-1})$ in the parameter region of interest,  the hierarchy ${p^2_*}/{\Lambda^2} \sim (\kappa_* \,{\chi_*}/{\Lambda})^2 \sim \kappa_*^2 \,\delta^2_{\rm H}$, with $\delta_{\rm H}$ given by Eq.~\eqref{deltaH}, remains parametrically suppressed even for $\Lambda \sim H_*$ ($\delta_{\rm H}\sim 1$).  Therefore, the tachyonically amplified gradients do not probe momenta comparable to the EFT cutoff,  and higher-derivative operators involving additional powers of $\nabla^2/\Lambda^2$ remain consistently subleading.  The instability thus develops entirely within the derivative regime where the truncated EFT expansion is self-consistent.

An important consequence of the tachyonic amplification for $\nu= {\cal O}(1)$ concerns the classical nature of the resulting fluctuations. This quantum-to-classical transition is a well-understood and generic feature of tachyonic instabilities and parametric amplification in non-trivial backgrounds.~\footnote{A detailed analysis can be found e.g.~in Refs.~\cite{Polarski:1995jg,Lesgourgues:1996jc,Kiefer:1998qe,Garcia-Bellido:2002fsq}, to which we refer the reader for a comprehensive discussion.} The essence of the argument can be summarized as follows. 
\begin{figure}
\centering
    \includegraphics[width=0.8\textwidth]{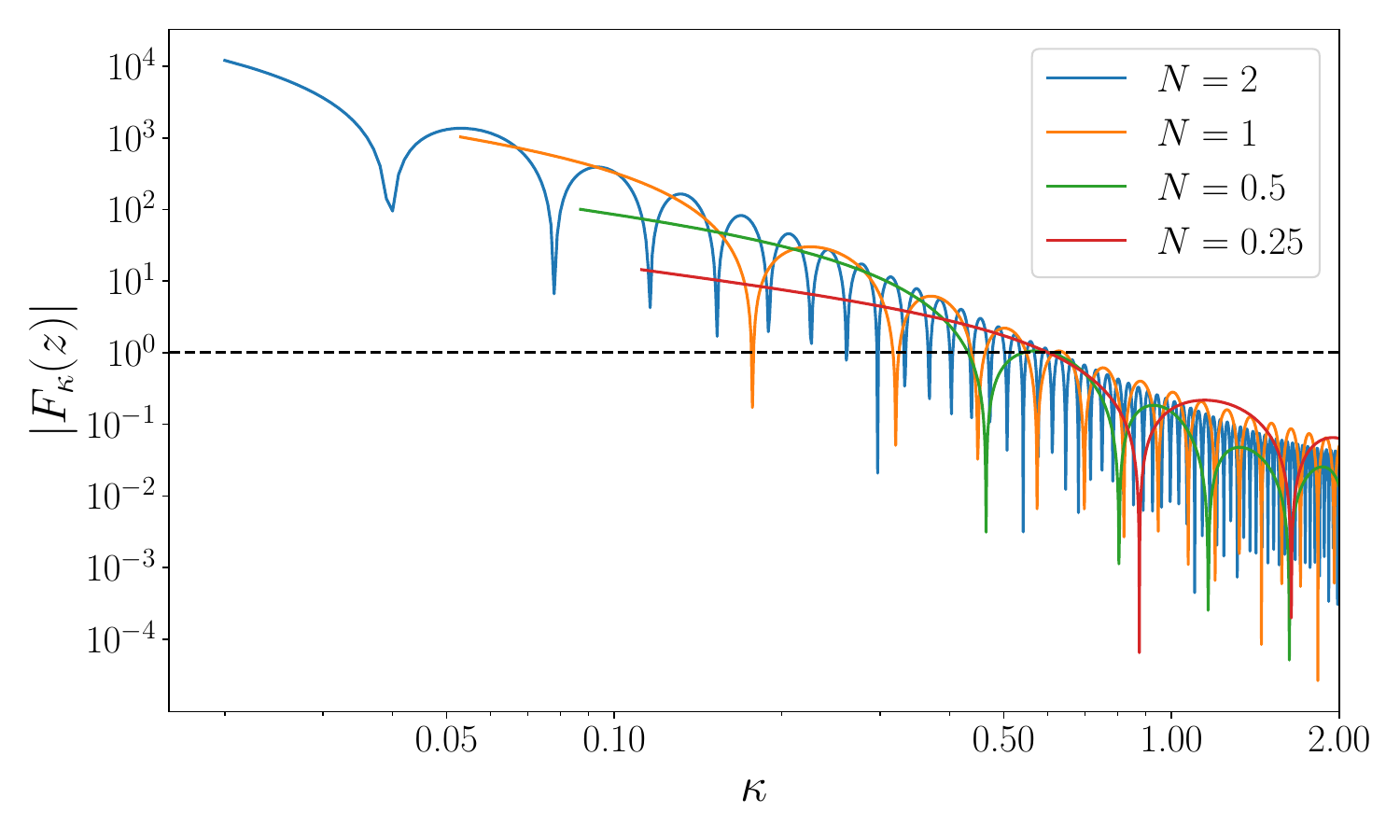}
  \caption{Evolution of the WKB phase $|F_\kappa(z)|$ at $N=0.25,\,0.5,\,1,$ and $2$ $e$-folds since the onset of RD for $\Lambda=H_*$, obtained by numerically solving the mode equation \eqref{eq:quantum_freq}. The horizontal black line marks $|F_\kappa|=1$. The rapid growth of $|F_\kappa|$ for IR modes signals strong squeezing and the onset of classical behavior for the amplified fluctuations.}
  \label{fig:WKB_phase}
\end{figure}
For each Fourier mode, the Heisenberg uncertainty relation in phase space takes the form
\begin{equation}
\Delta Y_\kappa^2\,\Delta \Pi_\kappa^2
= |F_\kappa(z)|^2+\frac{1}{4}\;\geq\; \frac{1}{4}\left|\left\langle\left[ Y_\kappa(z),\Pi_\kappa(z) \right]\right\rangle\right|^2\,,
\label{eq:classicalization_constraint}
\end{equation}
with $F_\kappa(z)$ the so-called WKB phase, defined as 
\begin{equation}
    F_{\kappa}(z) \equiv \frac{1}{2} \langle \Pi_{\vec{\kappa}}(z) Y_{\vec{\kappa}}^{\dagger}(z) + Y_{\vec{\kappa}}(z) \Pi_{\vec{\kappa}}^{\dagger}(z) \rangle = -\frac{i}{2} \left(g_{\kappa} f_{\kappa}^* - f_{\kappa} g_{\kappa}^*\right) = \text{Im}\{f_{\kappa}^* g_{\kappa}\}\,,
\end{equation}
and $g_{\kappa} \equiv i f_{\kappa}'$. When $|F_\kappa(z)|\gg1$, or equivalently when the anticommutator $\langle\{Y_\kappa,\Pi_\kappa^\dagger\}\rangle$ dominates over the commutator $\langle[Y_\kappa,\Pi_\kappa^\dagger]\rangle=\hbar$, the ambiguity associated with operator ordering becomes negligible. In this regime, each amplified quantum mode behaves indistinguishably from a classical stochastic variable.  As shown in Fig.~\ref{fig:WKB_phase}, the classicalization condition $|F_\kappa(z)|\gg1$ in the present scenario is satisfied for the vast majority of modes with $\kappa<1$, in agreement with the instability band in Eq.~\eqref{eq:momentum_interval}, since $\kappa_{\rm max}(z=0)=1$. As a result, the long-wavelength fluctuations generated during the tachyonic phase can be accurately described as realizations of a classical Gaussian random field. Quantum expectation values of observables can therefore be computed as ensemble averages over classical field configurations, with initial conditions drawn from the corresponding power spectrum. This will play an important role in the numerical analysis to be performed in Section \ref{sec:simulations}.

\section{Estimation of the dark matter relic density}\label{sec:relicdensity}

Building on the results obtained above, we estimate now the energy density injected into the spectator field sector at production and track its subsequent cosmological evolution. Our goal is to derive the parametric dependence of the present-day spectator abundance on the fundamental scales $H_*$, $M$, and $\Lambda$, and to clarify why these dependencies are largely insensitive to the detailed microphysics of the tachyonic phase.

\begin{figure}
  \centering
  \includegraphics[width=0.9\textwidth]{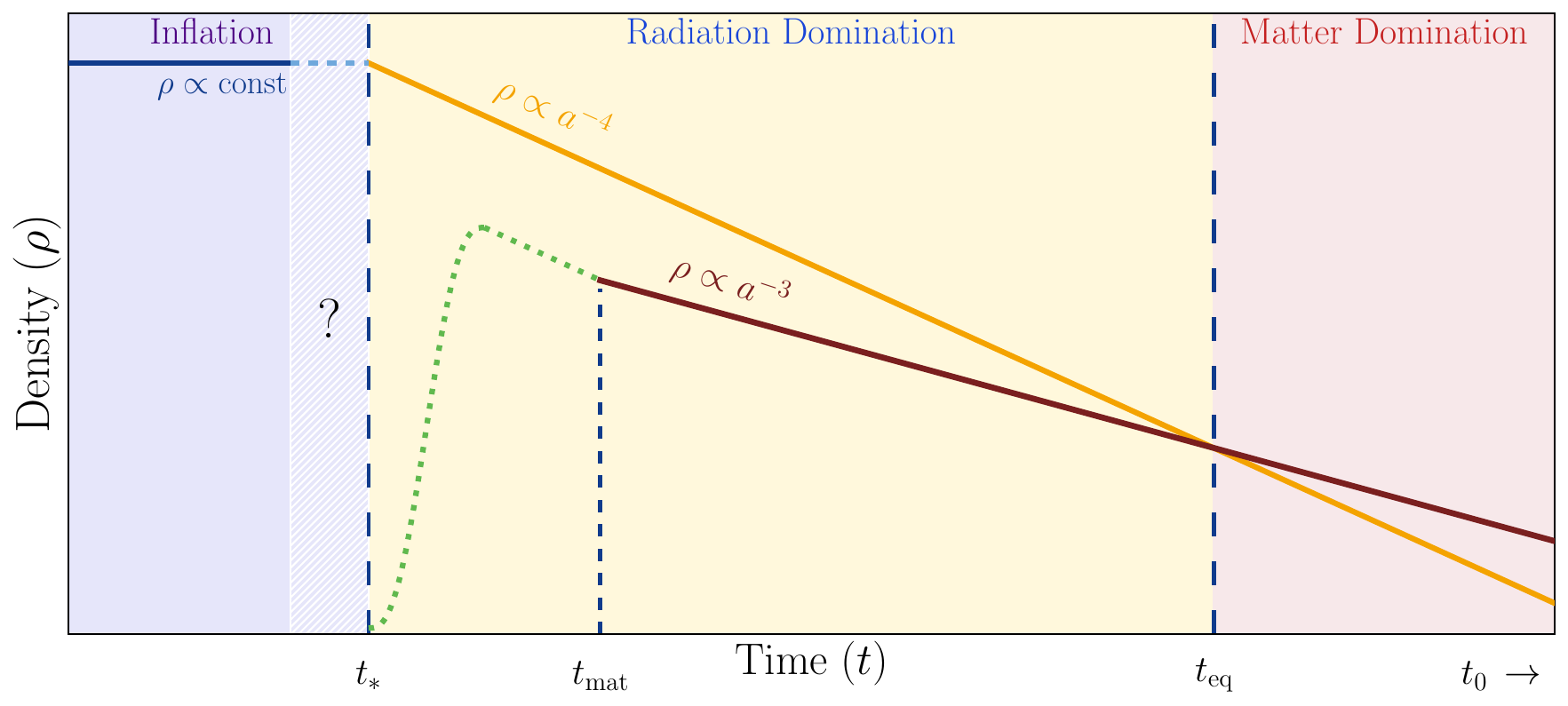}
  \caption{Schematic evolution of energy densities in the early Universe. The inflaton energy density (blue) is approximately constant until its end, while radiation (yellow) and matter (red) scale as $a^{-4}$ and $a^{-3}$, respectively. The green dotted curve illustrates the curvature-induced production of DM, initiated at the onset of RD ($t=t_\ast$) and terminating once the dark sector reaches matter-like behavior at $t=t_{\rm mat}$. The two components become equal at matter--radiation equality $t_{\rm eq}$, while $t_0$ denotes the present time.}
  \label{fig:scaling}
\end{figure}

As shown in the previous section, immediately after the tachyonic phase the amplified power spectrum $\Delta_\chi$ is sharply peaked around a characteristic physical momentum $p_* \sim \kappa_*\,\chi_*$, where numerically $\kappa_* \sim \mathcal O(10^{-2}\text{--}10^{-1})$. Consequently, for a broad and well--motivated region of parameter space satisfying $M \ll p_*$, the produced excitations are relativistic at the time of production. Their energy density is therefore dominated by gradient and kinetic contributions, while the bare mass term $M^2\chi^2$ is negligible in this regard. In this regime, the typical mode frequency is $\omega_k \simeq {k}/{a_*} \simeq \chi_*\,\kappa$,  and the energy density can be estimated as
\begin{equation}
\rho_{\chi,\mathrm{prod}}\sim \chi_*^4 \int \frac{d^3\kappa}{(2\pi)^3}\, \kappa^2 |f_\kappa|^2 =\frac{\chi_*^4}{2\pi^2}\,\mathcal G(\nu)\,,
\end{equation}
with 
\begin{equation}
\mathcal  G(\nu)\equiv \int_0^1 d\kappa\; \kappa^4\,|f_\kappa|^2 \simeq \int_0^1 d\kappa\; \kappa^3 \exp\!\left[\frac{4}{3}\,\mu_0^{3/2}(\kappa)\right]\, \mathcal P(\kappa,z_{\rm prod})
\end{equation}
a dimensionless growth factor encoding the efficiency of tachyonic amplification. Since the integral is dominated by the vicinity of the maximum $\kappa=\kappa_*$ of the spectrum, a crude saddle-point estimate gives
\begin{equation}
\mathcal G(\nu)\simeq \sqrt{\frac{2\pi}{|\frac43 q''(\kappa_*)|\nu}}\, 
\mathcal P_*\,\kappa_*^{3}\, \exp\!\left[\frac43q(\kappa_*)  \nu \right]  \simeq  2^{1/4}\sqrt{\frac{3\pi}{\,\nu}}\,\kappa_*^{3}\,(1-\kappa_*)^{1/4}\,
\mathcal P_*\,\exp\!\left[\alpha(\kappa_*)\,\nu\right]
\,,
\end{equation}
where we have taken into account Eq.~\eqref{mu0kone}, suitable for a scale $\kappa_*$ close to the instability threshold, $\kappa_*\lesssim 1$, and defined 
\begin{equation}
\mathcal  P_* \equiv \mathcal  P(\kappa_*,z_{\rm prod})=\mathcal{O}(1)\,,\qquad \alpha(\kappa_*)\equiv \frac{4\sqrt{2}}{9}\,(1-\kappa_*)^{3/2} 
\end{equation}
Normalizing the resulting energy density to the background one at the onset of RD, $\rho_{\rm R}(t_*)=3M_P^2H_*^2$, and using $\chi_*=\sqrt{48}\,H_*^2/\Lambda$, we obtain
\begin{equation}\label{rhoproduction} 
\frac{\rho_{\chi,\mathrm{prod}}}{\rho_R(t_*)}\sim \frac{384}{\pi^2}   \frac{H_*^6}{M_P^2\Lambda^4}\, \mathcal  G(\nu)\,.
\end{equation}
This result fixes the initial conditions for the post-instability evolution of the spectator sector. From this point onward, the relic spectator abundance depends only on the subsequent redshifting of the produced excitations, as schematically illustrated in Fig.~\ref{fig:scaling}. 
As long as $p_*>M$, the dark-sector component behaves effectively as a relativistic fluid with equation of state $w\simeq 1/3$, and its energy density redshifts as $\rho_\chi\propto a^{-4}$. Consequently, the ratio $\rho_\chi/\rho_{\rm R}$ remains approximately constant. However, at a later time $t=t_{\rm mat}$ the typical physical momentum redshifts down to the bare mass scale, $p(a_{\rm mat})\sim M$, with $a_{\rm mat}=a(t_{\rm mat})$. Since physical momenta redshift as $p(a)=p_*\,(a_*/a)$ with $p_*\sim\kappa_*\,\chi_*$, this condition implies
\begin{equation}\label{pstarscaling}
\frac{a_{\rm mat}}{a_*} \sim \frac{p_*}{M}\sim \sqrt{48}\,\kappa_*\,\left(\frac{H_*}{M}\right)\left(\frac{H_*}{\Lambda}\right)\,.
\end{equation}
At this point, the spectator field enters a regime of coherent oscillations and its energy density begins to scale as pressureless matter, $\rho_\chi\propto a^{-3}$, while the radiation background continues to dilute as $\rho_{\rm R}\propto a^{-4}$. As a result, the ratio $\rho_\chi/\rho_{\rm R}$ grows linearly with the scale factor until today. Using the observed radiation abundance $\Omega_{\rm R}^0=\rho_{\rm R}^0/(3M_P^2H_0^2)$ and the fact that $\rho_\chi/\rho_{\rm R}$ is conserved during RD while the excitations are relativistic, ${\rho_\chi(a_{\rm mat})}/{\rho_R(a_{\rm mat})}\simeq{\rho_{\chi,\mathrm{prod}}}/{\rho_R(t_*)}$, it then follows that the present-day abundance parameter satisfies
\begin{equation}
\frac{\Omega_{\rm DM}^0}{\Omega_R^0}=\frac{\rho_\chi(a_{\rm mat})}{\rho_R(a_{\rm mat})}\,\frac{a_0}{a_{\rm mat}}\simeq\frac{\rho_{\chi,\mathrm{prod}}}{\rho_R(t_*)}\,\frac{a_0}{a_*}\frac{a_*}{a_{\rm mat}}\,.
\end{equation}
The scale factor at the onset of the matter-like regime is fixed by the condition \eqref{pstarscaling}. Using in addition that $a_0/a_*= \left({\rho_R(t_*)}/{\rho_{R,0}}\right)^{1/4}$ with $\rho_R(t_*)=3M_P^2H_*^2$ and $\rho_{R,0}=3M_P^2H_0^2\Omega_R^0$, the relic abundance can be written in the compact form
\begin{equation}\label{eq:abundance}
\Omega_{\rm DM}^0\simeq(\Omega_R^0)^{3/4}\, \left(\frac{H_*}{H_0}\right)^{1/2}
\frac{\rho_{\chi,\mathrm{prod}}}{\rho_R(t_*)}\,
\frac{M}{p_*}\,, 
\end{equation}
which, taking into account \eqref{rhoproduction}, becomes
\begin{equation} \label{eq:Omega_intermediate}
\Omega_{\rm DM}^0\sim \mathcal C(\nu,\kappa_*)\,  \left( \frac{H_*}{M_P}\right)^{5/2} \,  \left(\frac{M}{H_*}\right)\, \nu^{5/2} \, \exp\,[\alpha(\kappa_*)\, \nu]\,,
\end{equation}
with the dimensionless coefficient  
\begin{equation}
\mathcal C(\nu,\kappa_*)\sim \frac{(\Omega_R^0/2)^{3/4}}{\sqrt{3} \pi^{3/2}}\, \left(\frac{M_P}{H_0} \right)^{1/2} \mathcal P_* \, \kappa_*^{2}\,
(1-\kappa_*)^{1/4} 
\end{equation}
collecting numerical factors and phase-space suppression information. The linear dependence of Eq.~\eqref{eq:Omega_intermediate} on $M/H_*$ reflects the kinematical nature of the radiation-to-matter transition, which is controlled by the redshifting of momenta. On the other hand, the strong $\nu$ dependence directly traces back to the curvature-induced origin of the tachyonic instability. For $\Lambda\sim H_*$ and $\kappa_*\sim 10^{-1}$, the coefficients $\alpha(\kappa_*)$ and ${\mathcal C(\nu,\kappa_*)}$ are naturally of order $0.5$ and $10^{24}$, respectively. 

Imposing that the present-day abundance of the spectator field reproduces the observed dark-matter density $\Omega_{\rm DM}^0h^2\simeq 0.12$ \cite{Planck2018}, allows the parametric scaling in Eq.~\eqref{eq:Omega_intermediate} to be translated into an estimate of the spectator mass. For $\Lambda\sim H_*$ and $\kappa_*=10^{-1}$, we find that masses ranging from $\mathcal O(10^2)\,\mathrm{GeV}$ down to the sub-GeV regime arise naturally for inflationary scales $H_*\sim10^{10}\text{--}10^{13}\,\mathrm{GeV}$. 

\section{Lattice simulations}\label{sec:simulations} 

The analytical considerations developed in the previous sections allowed us to identify the dominant physical scales controlling curvature-induced tachyonic production and to derive robust parametric estimates for the resulting DM abundance. However, these arguments necessarily rely on simplifying assumptions, such as the dominance of a single momentum scale or the approximate radiation-like redshifting of the produced excitations. While sufficient to establish the correct scaling with $H_*$, $M$, and $\Lambda$ in Eq.~\eqref{eq:Omega_intermediate}, such estimates cannot fully capture the stochastic nature of quantum fields in the early universe and the detailed post-tachyonic dynamics, including the redistribution of energy among kinetic, gradient, interaction, and potential components, the time-dependent evolution of the equation of state, or the precise onset of a matter-like behavior. To accurately follow this evolution and reliably connect the early-time instability to the late-time cosmological observables, a numerical treatment in real space is therefore required. This motivates the use of classical lattice simulations within the spectator field approximation, which provide a natural and controlled framework to evolve the system beyond the regime accessible to analytical estimates.

We perform 3+1 dimensional lattice simulations of the spectator field dynamics on a fixed FLRW background using the \texttt{$\mathcal{C}osmo\mathcal{L}attice$} library \cite{Figueroa:2020rrl,Figueroa:2021yhd}. This framework discretizes the equations of motion on a spatial lattice with periodic boundary conditions and evolves them using finite-difference schemes with controlled numerical accuracy. Rather than solving the continuum Klein--Gordon equation~\eqref{eq:KG_Y} directly, the field is evolved on a cubic lattice with $N_L^3$ points in comoving coordinates, while the background expansion is treated as an external input, in agreement with the spectator field approximation. This simplifying assumption holds well within the parameter ranges explored in our simulations, allowing us to safely neglect any backreaction from the DM field and retain the standard Friedmann equations. Specifically, we ensure that the peak energy density of the spectator field consistently remains below 0.1\% of the total background energy density at production.

The lattice spacing $\delta x$ and box size $L=N_L\delta x$ are chosen such that all relevant physical scales are well resolved. 
In particular, the IR and UV momentum cutoffs, $\kappa_{\rm IR}=2\pi/L$ and $\kappa_{\rm UV}=\sqrt{3}N_L\kappa_{\rm IR}/2$, are required to cover the region of peak amplification, namely $\mathcal{H}(z_{\rm sr})<\kappa<\nu^3\mathcal{H}^3(0)$, being $\mathcal{H}(z_{\rm sr})\ll \kappa_*$ (cf.~Eqs.~\eqref{eq:momentum_interval} and \eqref{kappastareq}).  
These requirements translate into a lower bound on the lattice resolution, 
$N_L > 2\nu^3\,\mathcal{H}^3(0)/(\sqrt{3}\mathcal{H}(z_{\rm sr}))$, 
which is satisfied in all simulations presented in this work. We have further verified that our results are insensitive to the lattice resolution and UV cutoff; details of these consistency tests are provided in Appendix~\ref{app:lattice_checks}.

Initial conditions are specified at the onset of RD, $t=t_\ast$, as a homogeneous background plus fluctuations, $\chi(\vec{x},t_\ast)=\bar\chi_\ast+\delta\chi_\ast(\vec{x})$. We set $\bar\chi_\ast=\bar\chi'_\ast=0$, while the fluctuations are drawn from a Gaussian random field whose power spectrum matches that of the vacuum initial conditions considered in Section~\ref{sec:quantum}. The procedure for generating these initial conditions is described in detail in Ref.~\cite{Figueroa:2021yhd}. In practice, we fix a single random seed across all simulations to facilitate direct comparison between different parameter choices, but we have checked that this does not affect macroscopic observables, as the system rapidly loses memory of the initial realization once the tachyonic instability sets in.

The lattice simulations provide access to the full spatial distribution of the field and its energy components at each time step. From these, we compute volume-averaged quantities such as the kinetic, gradient, interaction, and potential contributions to the total energy density, as well as the pressure. Given the dominance of long-wavelength fluctuations and the rapid loss of memory of initial conditions, we invoke ergodicity and identify ensemble averages with spatial averages over the lattice volume. 

\begin{figure}
      \centering
      \includegraphics[width=0.9\textwidth]{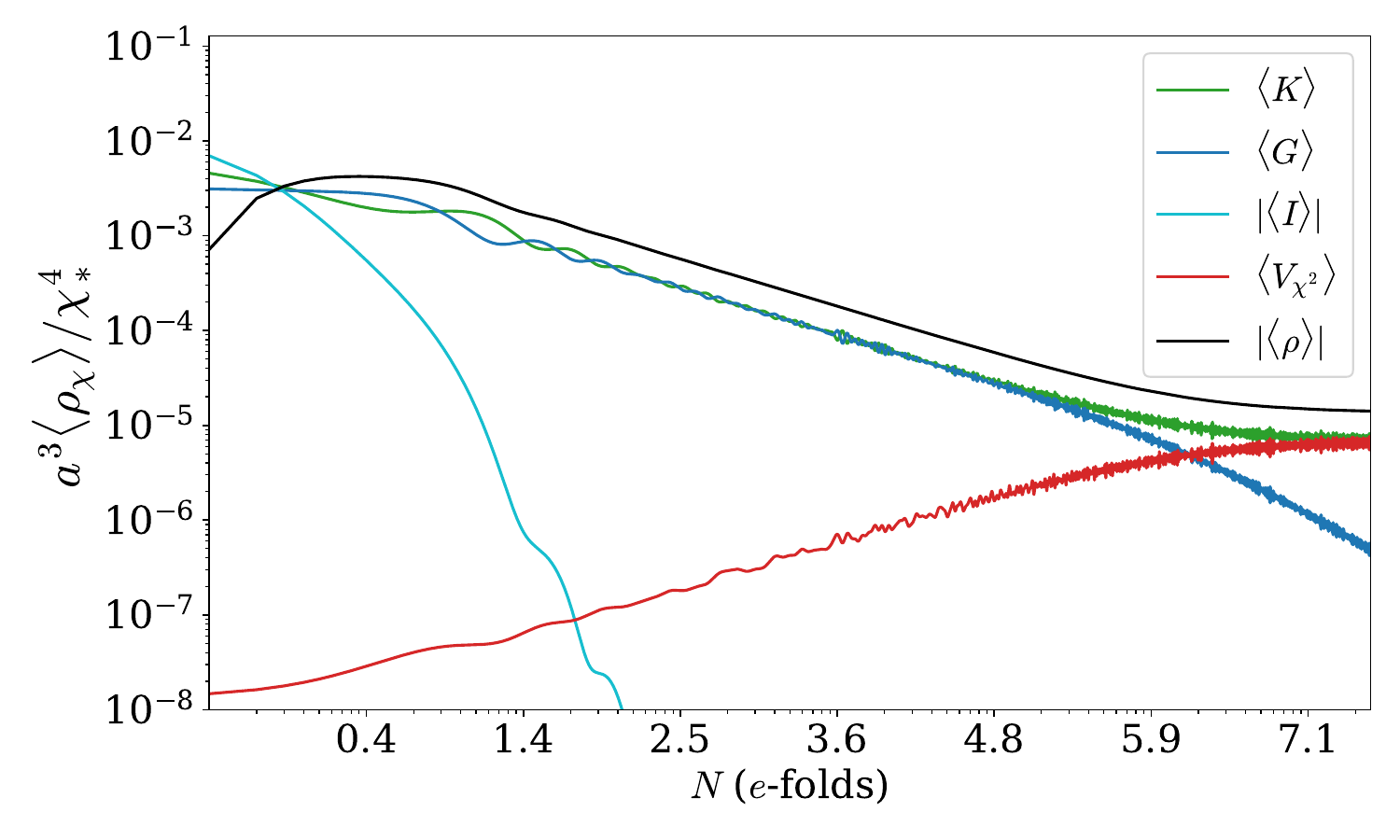}
      \caption{Late-time evolution of the volume-averaged energy-density components of the spectator field $\chi$ as a function of the number of $e$-folds since the onset of RD. The quantities $K$, $G$, $I$, and $V$ denote the kinetic, gradient, interaction, and potential contributions to the total energy density $\rho_\chi$, all normalized to $\chi_\ast^4$. The curves are rescaled by a factor $a^{3}$, so that horizontal lines correspond to matter-like scaling. The model parameters have been set to $H_\ast=10^{12}\,\mathrm{GeV}$, $\Lambda=5H_*$ and $M=10^{9}\,\mathrm{GeV}$, while the simulation parameters are given by $\kappa_{\rm IR}=0.12$, $\delta t=0.1$ and $N_L=128$.}
    \label{fig:en_dens_sim_long}
    \end{figure}

\begin{figure}
      \centering
      \includegraphics[width=0.9
    \textwidth]{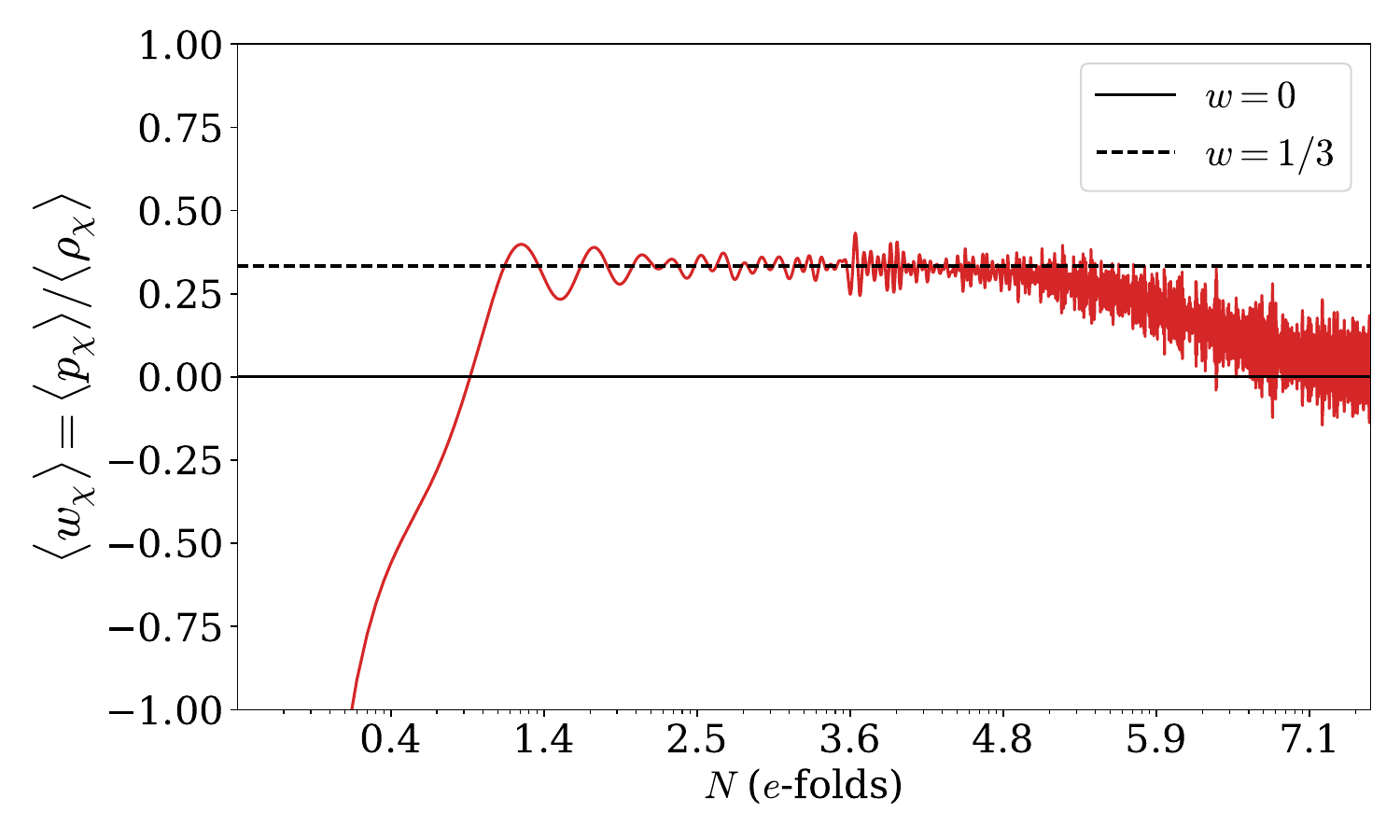}
      \caption{Time evolution in $e$-folds since the onset of RD of the volume-averaged equation-of-state parameter $w \equiv \langle p_\chi\rangle/\langle\rho_\chi\rangle$ extracted from lattice simulations. The dashed and solid black lines indicate the radiation-like ($w=1/3$) and matter-like ($w=0$) equations of state, respectively. After the tachyonic amplification phase, the field undergoes an extended intermediate stage in which it behaves effectively as a relativistic fluid, before smoothly transitioning to a pressureless matter regime once the bare mass $M$ becomes relevant. The numerical evolution of $w$ is derived from the same simulation in Figure \ref{fig:en_dens_sim_long} with $H_\ast=10^{12}\,\mathrm{GeV}$, $\Lambda=5H_*$, $M=10^{9}\,\mathrm{GeV}$ and $\kappa_{\rm IR}=0.12$, $\delta t=0.1$, $N_L=128$.}
    \label{fig:eos_long_plot}
    \end{figure}

An illustrative example of the information extracted from the simulations is shown in Fig.~\ref{fig:en_dens_sim_long}, where the evolution of the different energy-density components is displayed for a representative parameter choice. During the tachyonic phase, the interaction term associated with the curvature-induced coupling grows rapidly and temporarily dominates the energy budget. As the Universe expands, this contribution is quickly diluted, scaling as $H^4$ in RD, and becomes subdominant. Following this initial stage, the system enters a regime in which kinetic and gradient terms dominate, and the spectator field energy density scales approximately as $a^{-4}$, mimicking a radiation-like fluid. This behavior persists until the characteristic momenta of the field redshift below the bare mass scale $M$. At that point, the potential contribution becomes increasingly important and the system relaxes into a regime with $\langle K\rangle\simeq\langle V\rangle$, characteristic of coherent oscillations in a quadratic potential. The energy density then scales as $a^{-3}$, corresponding to a pressureless matter component. This change of behaviour is also reflected in the evolution of the averaged equation-of-state parameter in Fig.~\ref{fig:eos_long_plot}, which evolves from $w\simeq 1/3$ at early times to $w\simeq0$ at late times. Once the latter regime is reached, the comoving energy density of the spectator field is conserved and its subsequent evolution is entirely dictated by the expansion of the Universe. This allows us to reliably propagate the DM abundance to the present epoch.

 \begin{figure}
    \centering
  \begin{subfigure}[b]{0.49\textwidth}
    \centering
    \includegraphics[width=\textwidth]{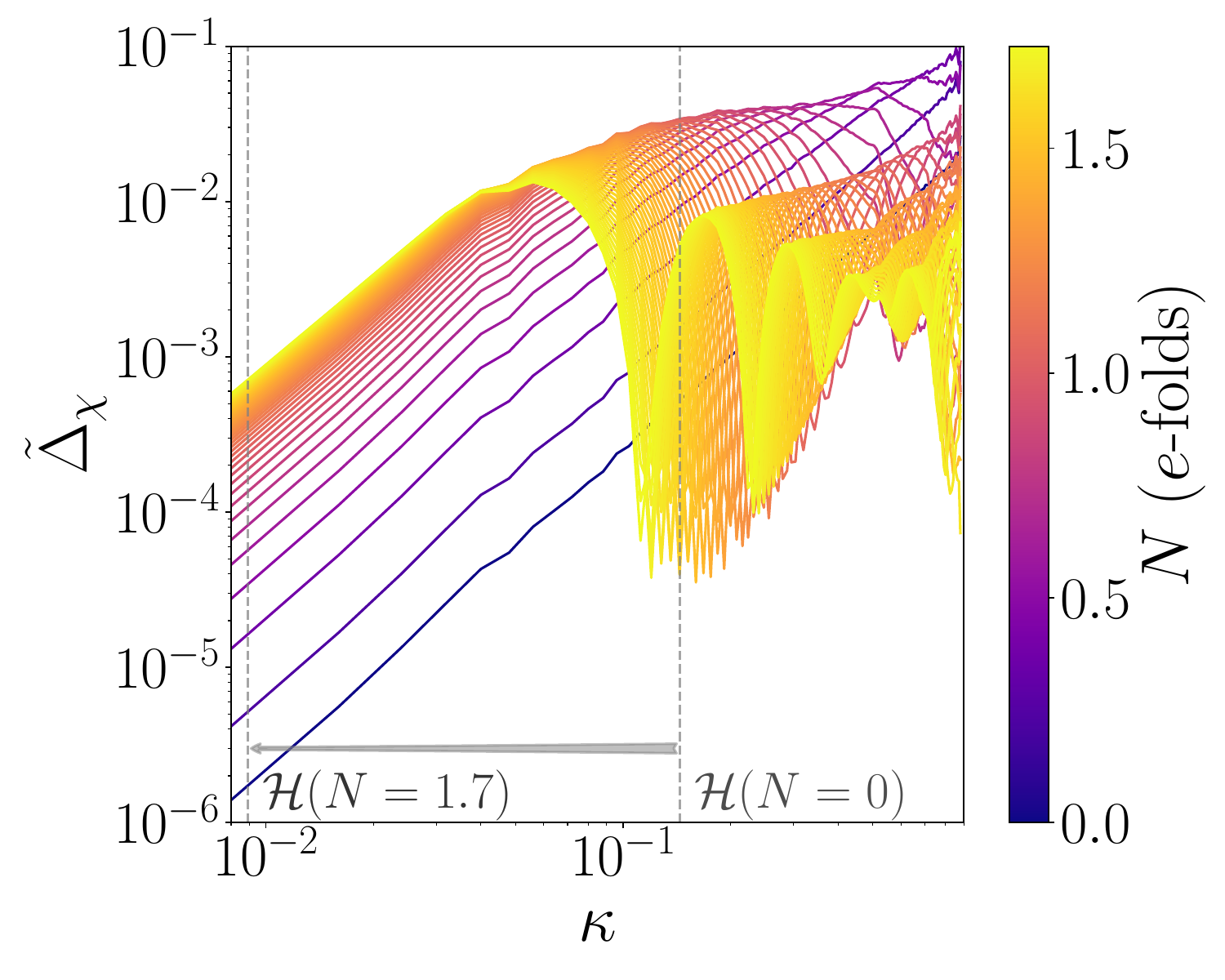}
    \label{fig:spectra_big}
  \end{subfigure}
    \hfill
  \begin{subfigure}[b]{0.49\textwidth}  
    \centering
    \includegraphics[width=\textwidth]{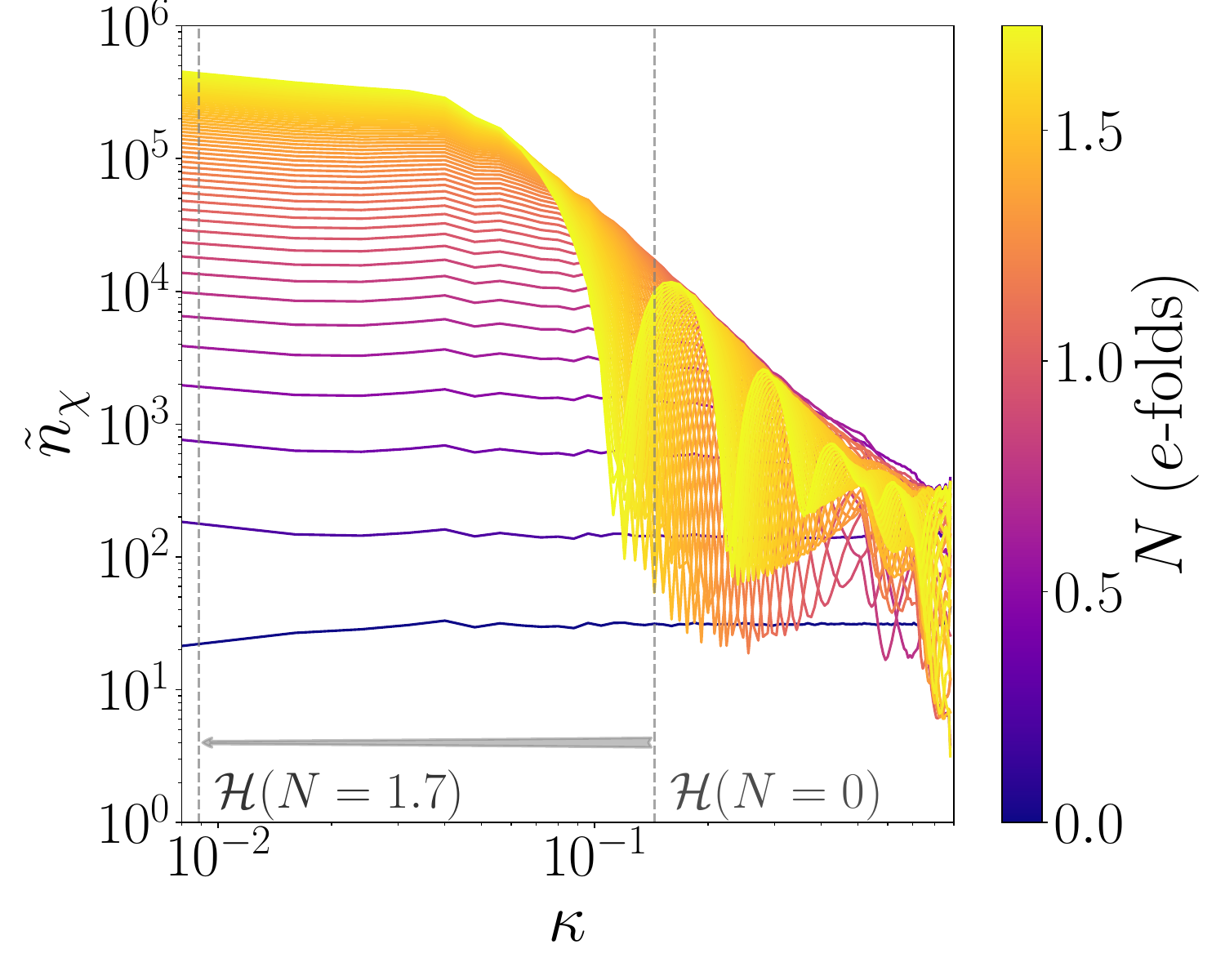}
    \label{fig:occup_num_big}
  \end{subfigure}
  \caption{Evolution of the normalized power spectrum of field fluctuations (left panel) and the corresponding lattice-defined occupation numbers (right panel), from $N=0$ $e$-folds until the symmetry restoration time $z_{\rm sr}$. These were obtained on a $128^3$ lattice, with $H_*=\Lambda=10^{12}\,\mathrm{GeV}$, $M=10^{8}\,\mathrm{GeV}$ and $\kappa_{\rm IR}=0.008$. The spectra are normalized to the scale $\chi_\ast$, and the color coding indicates different times during the evolution, expressed in $e$-folds since the onset of RD, with blue (yellow) curves corresponding to earlier (later) times. The grey dashed lines denote the comoving Hubble scale, from the beginning of tachyonic amplification up to $z_{\rm sr}$. A rapid and strong amplification of IR modes is observed, which dominates the dynamics and drives the subsequent evolution of the system.}
  \label{fig:spectrum}
\end{figure}

Another valuable piece of information that can be extracted from the lattice simulations is the power spectrum of the field fluctuations. The spectra shown in Fig.~\ref{fig:spectrum} confirm the qualitative behavior anticipated in the analytical treatment of Section~\ref{sec:quantum}. In particular, the tachyonic instability leads to a rapid and efficient amplification of IR modes, which come to dominate the energy density of the spectator field and drive the subsequent dynamics of the system. UV modes, by contrast, remain largely unaffected and retain their vacuum-like behavior throughout the full evolution. This IR-dominated structure is also reflected in the lattice-defined occupation number, constructed following the prescription of Ref.~\cite{Figueroa:2021yhd}. Due to the fact that the effective mass of the field is imaginary and the adiabatic particle interpretation breaks down, the notion of particle occupation number is not strictly well-defined during the tachyonic phase, but it nevertheless provides a useful diagnostic of mode amplification. In particular, it clearly captures the strong enhancement of long-wavelength modes relative to the UV sector and offers an intuitive measure of the redistribution of energy in momentum space.

To determine whether curvature-induced tachyonic production can account for the observed DM density within a physically viable region of parameter space, we perform an extensive numerical analysis. In total, we carried out approximately $200$ independent 3+1 classical lattice simulations, scanning several orders of magnitude in the three free parameters of the model, namely the DM bare mass $M$, the EFT cutoff scale $\Lambda$, and the Hubble rate at the end of inflation $H_*$. The region of parameter space that can be meaningfully explored is subject to several theoretical and observational constraints:
\begin{enumerate}
\item  \textit{DM mass bounds}: A lower bound on the DM mass is set by structure-formation considerations. $N$-body simulations and observations of ultra-faint dwarf galaxies imply $M \gtrsim 10^{-19}\,\mathrm{eV}$~\cite{PhysRevD.106.063517}, a bound that is further supported by analyses of the DM power spectrum~\cite{PhysRevLett.132.221004}. From above, the mass must remain smaller than the curvature-induced scale that controls the onset of the tachyonic instability, $M < \chi_* =\sqrt{48}H_*^2/\Lambda$, ensuring that the effective mass squared becomes negative at early times and that tachyonic amplification can occur.
\item  \textit{Hubble scale bounds}: The Hubble scale $H_*$ is identified with the Hubble rate at the end of inflation, assuming an instantaneous transition to RD. Successful Big Bang Nucleosynthesis requires $H_* \gtrsim 1\,\mathrm{MeV}$~\cite{PhysRevD.64.023508}, while the absence of a detected primordial tensor signal bounds the inflationary scale from above, $H_* \lesssim 10^{14}\,\mathrm{GeV}$~\cite{BICEP2:2018kqh}.
\item \textit{EFT restrictions}: The cutoff scale $\Lambda$ governing the validity of the gravitational EFT must lie between the Hubble scale and the Planck scale, $H_* \lesssim \Lambda \lesssim M_P$. For excessively large values of $\Lambda$, the tachyonic phase becomes extremely short-lived, suppressing both the amplification of fluctuations and their subsequent classicalization. To remain in a regime where the instability is active, numerically tractable with classical lattice simulations, and consistent with the EFT power counting discussed in Section \ref{sec:EFT}, we restrict our analysis to the range $\Lambda = \mathcal{O}(1\text{--}10)\,H_*$. The case $\Lambda \sim H_*$ (i.e. $\delta_H \sim 1$) is included only as a reference benchmark illustrating maximal tachyonic efficiency. 
Our main quantitative results and relic-density estimates rely instead on $\Lambda \gtrsim 2H_*$, where $\delta_H \leq 0.25$ and the curvature expansion remains manifestly hierarchical. 
In particular, for $\Lambda \gtrsim 5H_*\,, 10H_*$ one has respectively $\delta_H \lesssim 0.04,\,0.01$, ensuring that higher-curvature operators are parametrically suppressed by at least an order of magnitude relative to the leading dimension-six term in Eq.~\eqref{Snonrenorm}. In this regime, the tachyonic amplification does not probe momenta of order $\Lambda$ either, cf.~Section \ref{sec:quantum}. 

\item\textit{Matter-like behavior before equality:} The spectator field must enter the matter-like regime before standard matter--radiation equality. Since the transition occurs when the characteristic physical momentum redshifts to the bare mass, $p(a_{\rm mat})\simeq M$, this requires $a_{\rm mat}/a_*\simeq p_*/M < a_{\rm eq}/a_*$. During RD, $a_{\rm eq}/a_*=(H_*/H_{\rm eq})^{1/2}$, so in terms of number of $e$-folds the condition can be written as $N_{\rm mat}< N_{\rm eq}$, with $N_{\rm mat}\simeq\ln\!\left({p_*}/{M}\right)$ and $N_{\rm eq }=\frac12\ln\!\left({H_*}/{H_{\rm eq}}\right)$. 
\end{enumerate}

\begin{table}
\centering
\centering
\renewcommand{\arraystretch}{1.4}
\begin{tabular}{c c c c c c}
$\Lambda/H_*$ & $M$ (GeV)  & $H_\ast$ (GeV) & 
$\kappa_{\mathrm{IR}}$ &$\kappa_{\mathrm{UV}}$ &  $\delta x$\\
\hline
1  & $10^8$-$10^{13}$ & $10^{10}$-$10^{13}$ & 0.04  & 2.22 & 2.45\\[0.3em]
2  & $10^8$-$10^{13}$ & $10^{10}$-$10^{13}$ & 0.04  & 2.22 & 2.45\\[0.3em]
5  & $10^7$-$10^{12}$ & $10^{10}$-$10^{13}$ & 0.12  & 6.65 & 0.82\\[0.3em]
10 & $10^8$-$10^{11}$ & $10^{11}$-$10^{12}$ & 0.12  & 6.65 & 0.82\\[0.3em]
\hline
\end{tabular}
\caption{\label{tab:lattice_params} Summary of model and lattice parameters employed in the full set of 3+1 dimensional simulations used in the parameter scan. For each value of the EFT suppression scale $\Lambda/H_*$ we report the corresponding IR momentum cutoff $\kappa_{\mathrm{IR}} = 2\pi/L$ (in units of $a_* \chi_*$), chosen to properly resolve the tachyonic instability band. The UV cutoff $\kappa_{\mathrm{UV}} = \sqrt{3}\,N_L \kappa_{\mathrm{IR}}/2$ is sufficiently large in all cases to ensure that amplified modes are well contained within the dynamical range of the lattice. We also include the equivalent lattice spacing $\delta x$. All the simulations were performed on a $N_L=64$ lattice.}
\end{table}

We perform dedicated batches of simulations for four representative values of the ratio $\Lambda/H_*=\{1,2,5,10\}$. For each choice, we vary $M$ linearly and $H_*$ logarithmically, and extract the corresponding relic abundance using Eq.~\eqref{eq:abundance}. These datasets allow us both to identify viable regions of parameter space and to construct accurate interpolating functions that smoothly connect the simulation points. In order to optimize the computational efficiency of our scanning of the parameter space, we fix the number of lattice sites to $64^3$. This choice allows for a proper coverage of all the relevant dynamical scales of the system within a reasonable computational time. We have explicitly verified that this numerical setup is not affected by IR or UV artifacts (see discussion in Appendix \ref{app:lattice_checks}). Table \ref{tab:lattice_params} summarizes the parameter space we are exploring together with the corresponding optimal choices for lattice parameters.

\begin{figure}
  \centering
  \includegraphics[width=0.99
\textwidth]{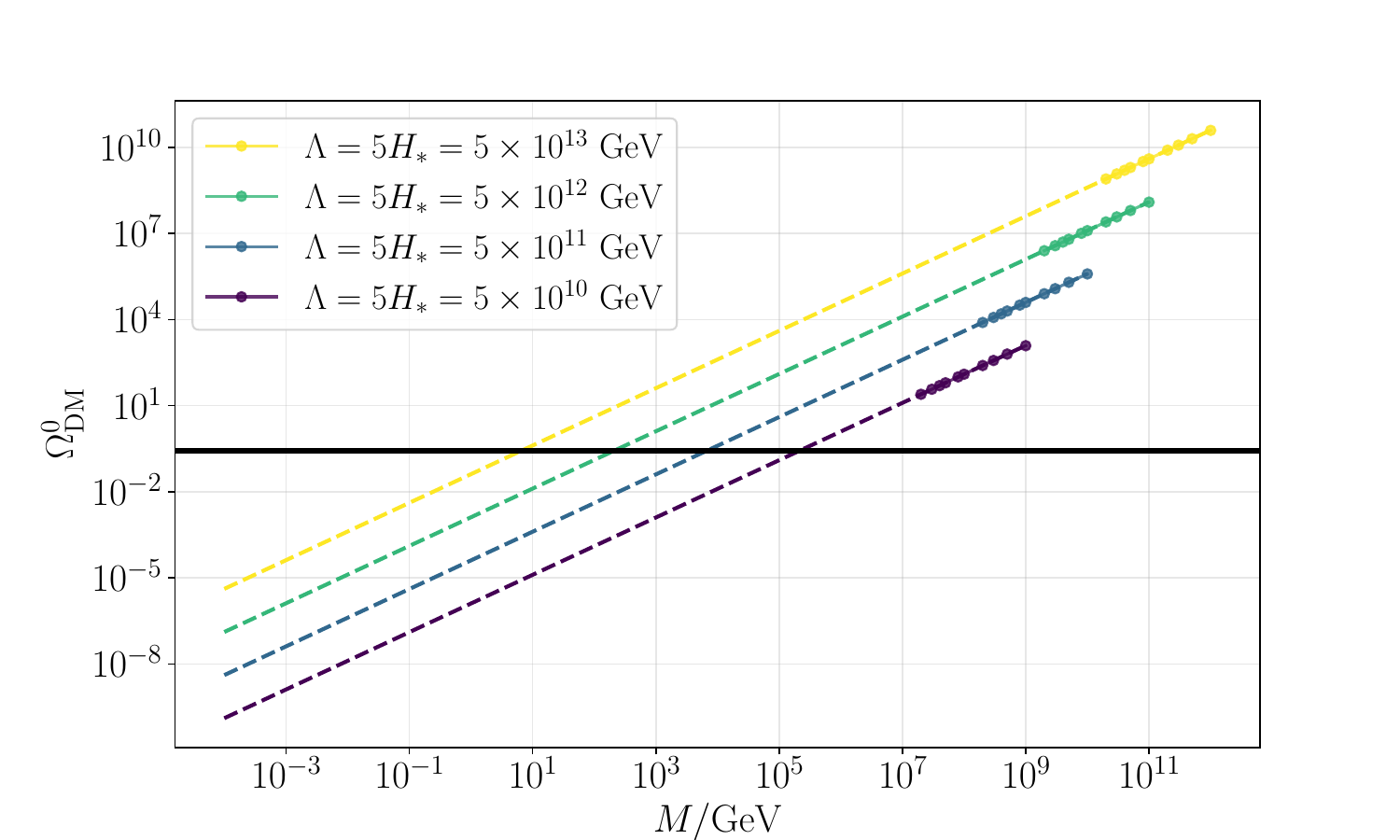}
  \caption{Present-day DM abundance $\Omega_{\rm DM}^0$ obtained from lattice simulations with $N_L=64$ lattice points and $\Lambda=5H_*$, shown as a function of the mass $M$ for different values of the Hubble scale at the end of inflation, $H_*=\{10^{13},10^{12},10^{11},10^{10}\}\,\mathrm{GeV}$. Each marker corresponds to an individual simulation, while the dashed curves represent interpolating fits to the numerical data, leading to the parametric expression in Eq.~\eqref{eq:interpolated_abundance_lambda}. The horizontal black line indicates the observed DM abundance inferred from CMB measurements, $\Omega_{\rm DM}^0h^2\simeq0.12$ \cite{Planck2018}.}
\label{fig:abundance_scan}
\end{figure}

\begin{figure}
  \centering
  \includegraphics[width=0.9\textwidth]{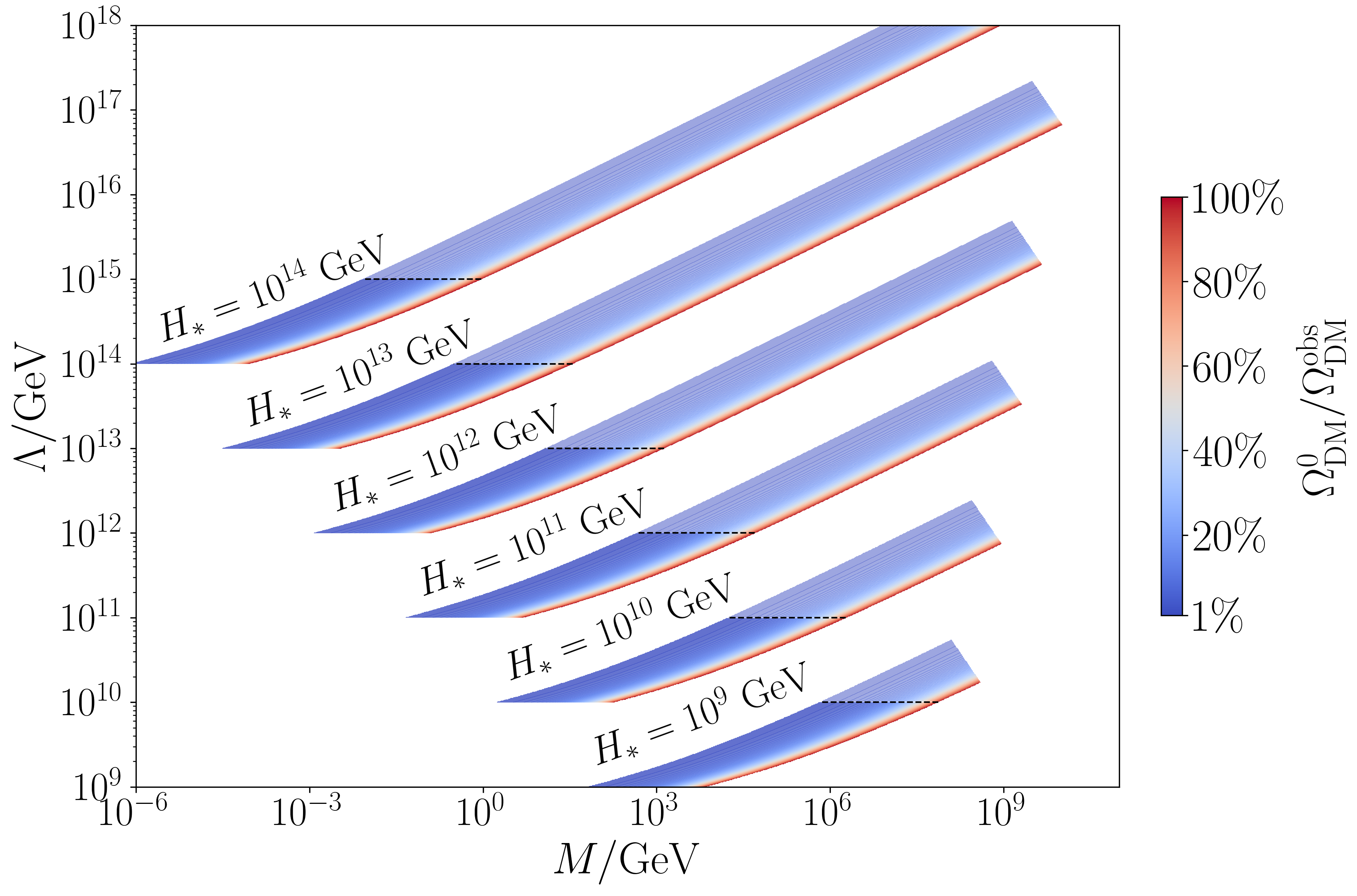}
  \caption{Exploration of the viable parameter space in the $(M,\Lambda,H_*)$ plane using the interpolated relic abundance formula in Eq.~\eqref{eq:interpolated_abundance_lambda}. For each fixed value of the Hubble scale at the end of inflation, $H_*$, the color scale indicates the predicted present-day DM abundance as a fraction of the observed value, with \textit{red} corresponding to $\Omega_{\rm DM}^0/\Omega_{\rm DM}^{\rm obs}=1$ and \textit{blue} to $\Omega_{\rm DM}^0/\Omega_{\rm DM}^{\rm obs}=10^{-2}$. For each $H_*$, points below the black dashed line correspond to the range spanned by the simulations $\Lambda=(1$–$10)\,H_*$. The points above this line correspond to an extrapolation beyond the strict validity region of our fit.}
\label{fig:final_scan}
\end{figure}

As a first illustration, Fig.~\ref{fig:abundance_scan} displays the present-day DM abundance $\Omega_{\rm DM}^0$ as a function of the particle mass $M$ for $\Lambda=5H_*$, and four representative inflationary scales $H_*=\{10^{13},10^{12},10^{11},10^{10}\}\,\mathrm{GeV}$. Each marker corresponds to an individual lattice simulation, while dashed lines provide smooth interpolations of the numerical results. The solid black line indicates the observed value of the present-day DM abundance $\Omega_{\rm DM}^0h^2\simeq0.12$~\cite{Planck2018}. Smaller values of $M$ lead to longer-lived tachyonic phases, hence the constraint on $\Omega_{\rm DM}$ is satisfied for higher inflationary scales.
Remarkably, the typical masses selected by this mechanism are generically much lighter than those arising in other gravitational DM production scenarios \cite{Kolb:2023ydq}. For $\Lambda=5H_*$, the correct relic density is obtained for $M\sim{\cal O}(10 \; \mathrm{  GeV})\!-\!{\cal O}(10^5\;\mathrm{ GeV})$ as $H_*$ varies from $10^{13}$ to $10^{10}\,\mathrm{GeV}$. Since the present-day abundance is fixed, $\Omega_{\rm DM}^0\propto M n$, this implies a parametrically larger number density $n$ at production compared to non-adiabatic gravitational mechanisms \cite{Cembranos:2019qlm,Cembranos:2023urh}, reflecting the intrinsic efficiency of curvature-induced tachyonic amplification, which generates exponentially large occupation numbers in IR modes.

Combining the simulation sets with $\Lambda/H_*=\{1,2,5,10\}$, we find that the relic DM abundance over the explored parameter space is accurately captured by the fitting formula
\begin{equation} \label{eq:interpolated_abundance_lambda}
\Omega_{\rm DM}^0(H_*, M, \Lambda)\sim 10^{a}\,  \left( \frac{H_*}{M_P}\right)^{b} \,  \left(\frac{M}{H_*}\right)^{c}\, \nu^{d} \, \exp\,[ \alpha_{n}\nu]  \,, 
\end{equation}
with 
\begin{equation}
a=24.31  \,, \qquad b=2.50  \,, \qquad  c=0.97 \,,  \qquad  d=2.86\,,  \qquad \alpha_{n}=0.38 \,,
\end{equation}
which is valid for $\Lambda \sim H_*$ up to $\Lambda \sim 10H_*$. This expression provides a compact numerical representation of the simulation results, allowing for an efficient exploration of the parameter space without the need for additional direct simulations and, most importantly, offering a non-trivial validation of the robustness of our earlier analytical estimates. In particular, both the overall normalization and the parametric dependence of Eq.~\eqref{eq:interpolated_abundance_lambda} are in good agreement with the estimate in Eq.~\eqref{eq:Omega_intermediate}, indicating that the latter already captures the essential physics of the problem despite its many simplifying assumptions.

Using Eq.~\eqref{eq:interpolated_abundance_lambda}, we map the allowed region in the $(M,\Lambda,H_*)$ parameter space, as shown in Fig.~\ref{fig:final_scan}. For each fixed value of $H_*$, the color gradient indicates the predicted DM abundance as a fraction of the observed value. The parameter space is bounded by the EFT validity condition $H_*<\Lambda<M_P$ and by the requirement $M<\chi_*$. As $H_*$ decreases, the viable region rapidly shrinks. In particular, for inflationary scales $H_* \lesssim 10^6\,\mathrm{GeV}$, the mechanism fails to reproduce the observed DM abundance, effectively excluding that regime.

Taken together, these results demonstrate that curvature-induced tachyonic production provides a robust and predictive mechanism for gravitational DM generation across a wide and well-motivated region of parameter space.

\section{Conclusions and outlook }\label{sec:discussion}

In this work we have introduced and systematically explored a novel mechanism for the gravitational production of DM, rooted in curvature-induced tachyonic instabilities operating after inflation. The central observation is that sharp transitions in the cosmic equation of state—most notably the transition from inflation to a decelerating expansion—can dynamically trigger a temporary sign flip in the effective mass of a spectator field coupled to curvature invariants. When this occurs, the dark sector undergoes a short-lived phase of spontaneous symmetry breaking entirely driven by gravitational dynamics, leading to an explosive amplification of IR modes and efficient non-thermal particle production.

Working within a controlled EFT of gravity, we have shown that this phenomenon is generic for a broad class of quadratic curvature couplings. Focusing on the Gauss--Bonnet invariant as a theoretically distinguished benchmark, we demonstrated that the emergence of a tachyonic instability is a robust consequence of the background expansion history rather than the result of fine-tuned parameter cancellations. This choice also preserves second-order gravitational field equations and avoids the introduction of additional propagating degrees of freedom, allowing the tachyonic dynamics to be cleanly isolated as a property of the spectator sector evolving on an otherwise standard Einsteinian background.

We combined analytical arguments with 3+1 classical lattice simulations in a fixed FLRW background to follow the evolution of the spectator field across the tachyonic phase and into the late-time regime relevant for observational cosmology. Analytical considerations clarified the structure of the instability band, the dominance of IR modes, and the conditions under which tachyonically amplified fluctuations classicalize. Lattice simulations then provided a complete and quantitative description of the post-instability dynamics, including the redistribution of energy among kinetic, gradient, and potential components, the evolution of the equation of state, and the emergence of a conserved comoving DM density.

A key outcome of our analysis is that curvature-produced DM naturally exhibits a radiation-to-matter transition. Shortly after production, the energy density is dominated by relativistic excitations, which redshift approximately as radiation. At later times, i.e. once the bare mass becomes dominant, the system settles into coherent oscillations redshifting as pressureless matter. The transition between the two scaling behaviors is clearly tracked in our lattice simulations and it plays a crucial role in determining the final DM relic abundance. This dynamical change in the DM field's properties highlights the importance of following the full evolution of the equation-of-state rather than assuming an instantaneous matter-like behavior.

Our extensive scanning with lattice simulations of the parameter space spanned by the DM mass, the EFT cutoff, and the inflationary Hubble scale allowed us to thoroughly study the DM production process. We showed that curvature-induced tachyonic production can robustly reproduce the observed DM abundance over a wide and well-motivated region of parameter space. The lattice-derived numerical results are accurately captured by a simple parametric fit for the DM abundance, in excellent agreement with our analytical expectations. Moreover, the fitting formula we provided enables a lattice-independent application of our findings, thus facilitating their use in broader phenomenological studies.

The curvature-induced production mechanism also implies that the resulting DM population is genuinely non-thermal. Since the spectator field is assumed to have negligible couplings both to the visible sector and to itself, the amplified fluctuations do not undergo kinetic equilibration at any stage of their evolution. The momentum distribution generated during the tachyonic phase is therefore preserved up to cosmological redshifting, and the subsequent transition to matter-like behavior reflects only the dilution of physical momenta as the Universe expands, rather than a thermalization process. A direct consequence of this non-thermal origin is the presence of a residual velocity dispersion, whose magnitude is set by the characteristic momentum scale imprinted during tachyonic amplification. While the IR-dominated nature of the spectrum naturally suppresses large momenta, the ratio between the typical physical momentum at production and the DM mass controls the duration of the relativistic phase and, consequently, the free-streaming length. Regions of parameter space in which the DM remains relativistic for an extended period may therefore exhibit a suppression of power on small cosmological scales, analogous to warm DM scenarios. Conversely, when the bare mass becomes dominant sufficiently early, the model smoothly approaches cold DM behavior. A detailed assessment of the resulting transfer function and its confrontation with small-scale structure constraints, such as Lyman-$\alpha$ forest observations, is left for future work.

Beyond establishing viability, our framework opens several promising directions for future work. While we have focused on a real scalar spectator field undergoing the spontaneous breaking of a $\mathbb{Z}_2$ symmetry, the underlying mechanism is considerably more general and is expected to extend to theories with richer global symmetries, such as $\mathbb{Z}_3$ or global $U(1)$ symmetries, as commonly considered in DM model building~\cite{Ma:2007gq,Belanger:2012zr,Bernal:2015bla}. Any DM field whose effective mass receives curvature-dependent contributions may undergo a similar tachyonic phase during transitions to decelerating expansion. This suggests that curvature-induced symmetry breaking could occur for more general global symmetry structures, including larger discrete groups or continuous symmetries, as well as in multi-field settings where curvature effects select preferred directions in field space.

Extensions to other types of DM candidates are also conceivable, albeit more model dependent. Fermionic DM could be produced indirectly through Yukawa couplings to scalar fields undergoing curvature-driven symmetry breaking à la \textit{Combined Preheating}  \cite{Garcia-Bellido:2008ycs,Rubio:2015zia,Repond:2016sol,Fan:2021otj,Mansfield:2023sqp}, while vector or gauge-field realizations may be viable provided that curvature couplings are embedded into a fully gauge-invariant framework, for instance via Higgs or Stueckelberg mechanisms or higher-dimensional operators constructed from field strengths \cite{Alonso-Alvarez:2019ixv,Ozsoy:2023gnl,Cembranos:2023qph,Capanelli:2024rlk,Grzadkowski:2024oaf,DeFelice:2025ykh}. In such scenarios, the qualitative picture of curvature-triggered amplification followed by classicalization may persist, although the detailed dynamics and phenomenology could differ substantially.

A particularly interesting avenue concerns the interplay between curvature-induced tachyonic production and more realistic reheating scenarios. While our analysis assumed an effectively instantaneous transition to RD, prolonged reheating or preheating phases could imprint additional structure on curvature invariants, potentially leading to multiple tachyonic episodes or parametric resonance effects. Although such resonances are typically less efficient than tachyonic amplification, they may leave observable features in specific inflationary realizations and warrant further investigation.

Finally, the violent, out-of-equilibrium dynamics associated with curvature-driven symmetry breaking may also source secondary cosmological observables. In particular, the rapid growth of long-wavelength fluctuations and their subsequent relaxation generate anisotropic stresses that can act as a source of stochastic gravitational waves. While a quantitative prediction of the resulting spectrum requires extending the lattice simulations to include metric perturbations, the characteristic frequency range is set by the Hubble scale at production, suggesting potential sensitivity to high-frequency gravitational-wave detectors. Exploring these signatures would provide an independent probe of gravitational DM production mechanisms and further constrain the viable parameter space.

\section*{Acknowledgments}

The numerical 3+1 classical lattice simulations in this paper were carried out with the support of the Centro Nacional de Computa\c c\~ao Avançada (CNCA) funded by the Funda\c c\~ao para a Ci\^encia e a Tecnologia (FCT). The authors further acknowledge the computing resources and technical assistance provided by CENTRA/IST. Early simulations were performed on the “Baltasar-Sete-Sóis” cluster, supported by the H2020 ERC Consolidator Grant "Matter and strong field gravity: New frontiers in Einstein's theory" grant agreement no. MaGRaTh-646597. G.~L. acknowledges support from the Alexander von Humboldt Foundation. T.~M. gratefully acknowledges the hospitality and support of Ana Mour\~ao and the COSTAR group at IST, where some of this research was conducted. T.~M. also thanks Matteo Piani for helpful discussions. J.~R. is supported by a Ram\'on y Cajal contract of the Spanish Ministry of Science and Innovation with Ref.~RYC2020-028870-I. This research was further supported by the project PID2022-139841NB-I00 of MICIU/AEI/10.13039/501100011033 and FEDER, UE, and the 2025 Leonardo Grant for Scientific Research and Cultural Creation from the BBVA Foundation. The BBVA Foundation is not responsible for the opinions, comments, and content included in the project and/or its resulting outcomes, which are the sole and exclusive responsibility of the authors.

\appendix

\section{Lattice simulations consistency tests} \label{app:lattice_checks}

In this Appendix, we present a set of consistency tests demonstrating that the results presented in the main body of this work are insensitive to the lattice resolution and to the choice of IR and UV momentum cutoffs. To this end, we compare simulations performed with different grid sizes  $N_L$ and box lengths $L$, keeping the physical parameters  $(H_*,M,\Lambda)$ fixed. For each configuration we monitor the evolution of the total spectator energy density $\rho_\chi$.
\begin{figure}[!t]
    \centering
  \begin{subfigure}
  {0.8\textwidth}
    \centering
    \includegraphics[width=\textwidth]{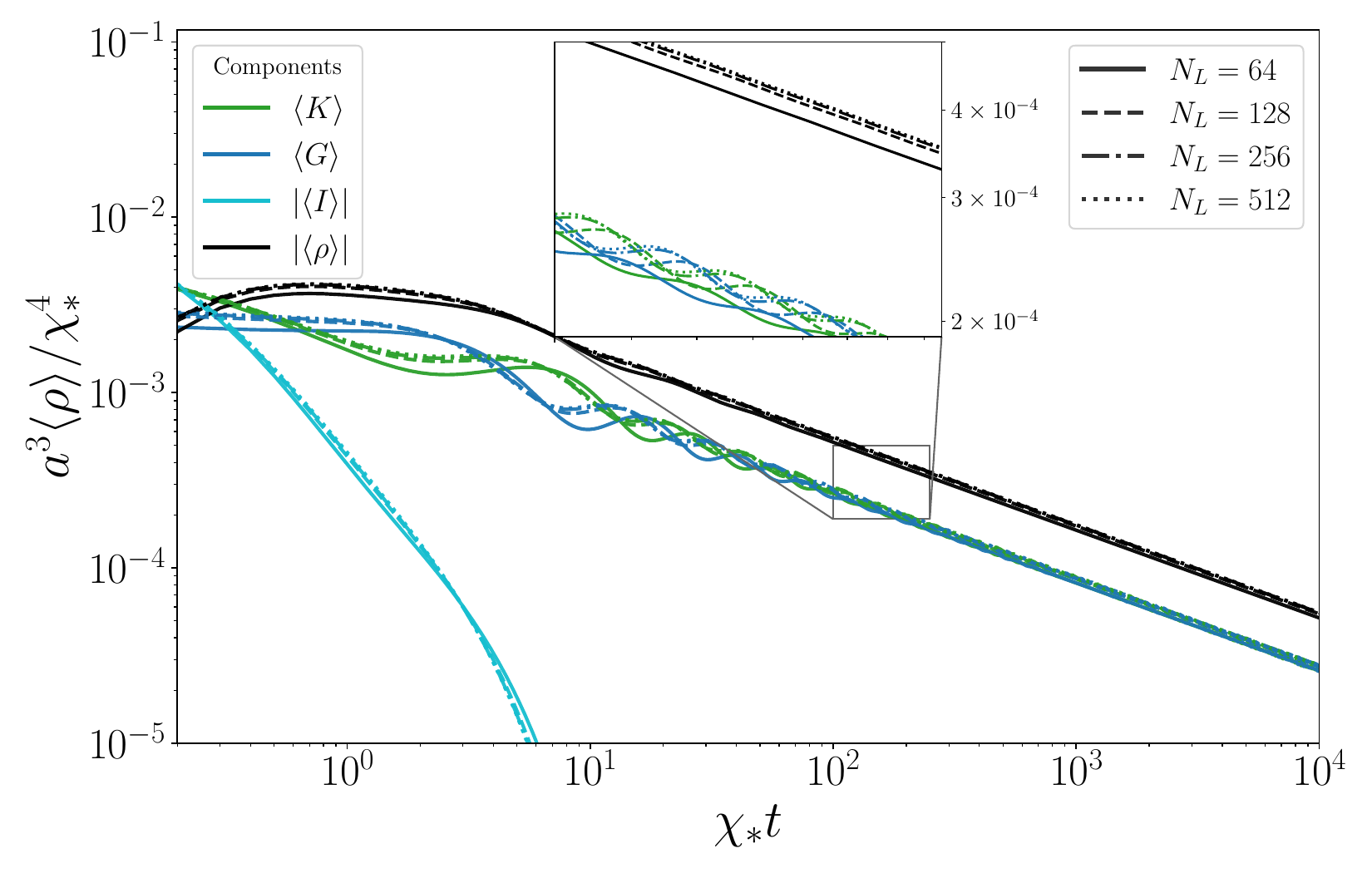}
    \caption{Fixed $\kappa_{\rm IR}$}
    \label{fig:energies_test_kIR}
  \end{subfigure}
    \hfill
  \begin{subfigure}{0.8\textwidth}  
    \centering
    \includegraphics[width=\textwidth]{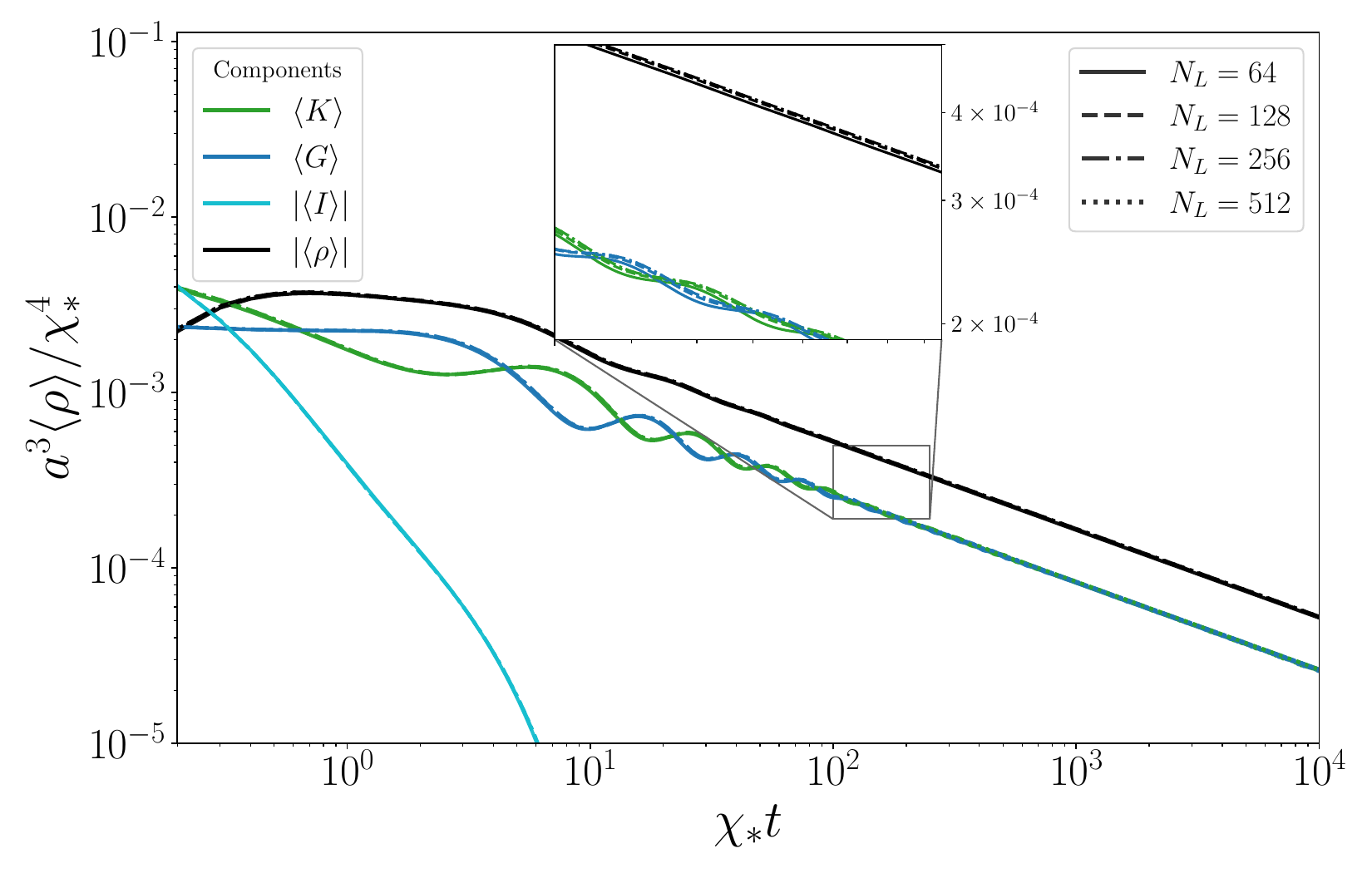}
    \caption{Fixed $\kappa_{\rm UV}$}
    \label{fig:energies_test_kUV}
  \end{subfigure}
  \caption{Evolution of the volume-averaged spectator energy density components - kinetic ($\langle K\rangle$), gradient ($\langle G\rangle$), interaction ($|\langle I\rangle|$) and the full quantity $|\langle\rho_\chi\rangle|$, with fixed $\kappa_{\rm IR}=0.04$ (\ref{fig:energies_test_kIR}) and fixed $\kappa_{\rm UV}=2.22$ (\ref{fig:energies_test_kUV}) for different number of lattice sites $N_L$, keeping $(H_*,M,\Lambda)=(10^{12},10^3,5\times10^{12})$ GeV fixed. Note that the potential contribution $V(\chi)=1/2 M^2\chi^2$ is not shown, because, in the time range considered, it remains negligibly small (of order of $10^{-20}$). All curves agree within numerical accuracy during the tachyonic phase and the subsequent radiation-like regime ($\rho_\chi \propto a^{-4}$).}
  \label{fig:resolution_energies}
\end{figure}

Fig.~\ref{fig:resolution_energies} shows the time evolution of the volume-averaged kinetic $K$, gradient $G$, interaction $I$ and total energy density $\rho_\chi$ for different IR and UV resolutions. In the upper panel four simulations are performed with increasing numbers of lattice sites while keeping $\kappa_{\rm IR}$ constant. The resulting energy densities are effectively identical for $N_L\geq128$, while the case $N_L=64$ shows a minimal offset of roughly 5\% at the onset of the radiation-like behavior. In the lower panel $\kappa_{\rm UV}$ is kept constant for different values of $N_L$. In this case, the curves are indistinguishable within numerical accuracy throughout 
the tachyonic phase and the subsequent radiation-like regime. These tests confirm that the main tachyonic production takes place within an IR instability band that is well captured by the numerical framework described in Section \ref{sec:simulations}. In particular, the peak lattice-averaged energy density reached during the instability and the late-time scaling coincide across all lattice configurations.

\FloatBarrier
\bibliographystyle{JHEP}
\bibliography{biblio}

\end{document}